\documentclass[english,prd,nofootinbib,preprintnumbers,twocolumn,showpacs]{revtex4}
\usepackage[latin1]{inputenc}
\usepackage{amsmath}
\usepackage{amsfonts}
\usepackage{appendix}
\usepackage{amssymb}
\usepackage{epsfig}
\usepackage{graphics,psfrag,rotating}
\usepackage{graphicx}
\usepackage{dcolumn}
\usepackage{bm}
\usepackage{epstopdf}
\usepackage{color}
\usepackage[usenames,dvipsnames,svgnames]{xcolor}
\usepackage[colorlinks=true,
            linkcolor=red,
            urlcolor=gray,
            citecolor=blue]{hyperref}
\usepackage{hyperref}
\usepackage[T1]{fontenc}
\usepackage{multirow}
\usepackage{float}
\usepackage{cancel}
\usepackage{hyperref}
\usepackage{adjustbox}
\usepackage{subfigure}

\usepackage{enumitem}

\def\3nab{\tilde{\nabla}}

\def\be {\begin{equation}}
\def\ee {\end{equation}}
\def\ba {\begin{align}}
\def\ea {\end{align}}

\def\bc {\begin{center}}
\def\ec {\end{center}}
\def\case#1/#2{\frac{#1}{#2}}

\newcommand{\bver}{\begin{verbatim}}
\newcommand{\ever}{\end{verbatim}}

\newcommand{\bea}{\begin{eqnarray}}
\newcommand{\eea}{\end{eqnarray}}
\newcommand{\beaa}{\begin{eqnarray*}}
\newcommand{\eeaa}{\end{eqnarray*}}

\newcommand{\lsim}   {\mathrel{\mathop{\kern 0pt \rlap
  {\raise.2ex\hbox{$<$}}}
  \lower.9ex\hbox{\kern-.190em $\sim$}}}
\newcommand{\gsim}   {\mathrel{\mathop{\kern 0pt \rlap
  {\raise.2ex\hbox{$>$}}}
  \lower.9ex\hbox{\kern-.190em $\sim$}}}
\def\case#1/#2{\textstyle\frac{#1}{#2}}

\begin{document}

\title{The Fisher gAlaxy suRvey cOde ($\texttt{FARO}$)}

\author{
Miguel Aparicio Resco 
}
\email{migueapa@ucm.es}
\affiliation{Departamento de F\'{\i}sica Te\'orica and Instituto de F\'{\i}sica de Part\'{\i}culas y del Cosmos (IPARCOS), Universidad Complutense de Madrid, 28040 
Madrid, Spain}
\author{Antonio L.\ Maroto}
\email{maroto@ucm.es}
\affiliation{Departamento de F\'{\i}sica Te\'orica and Instituto de F\'{\i}sica de Part\'{\i}culas y del Cosmos (IPARCOS), Universidad Complutense de Madrid, 28040 
Madrid, Spain}




\begin{abstract} 

The Fisher gAlaxy suRvey cOde ($\texttt{FARO}$) is a new public Python code that computes the Fisher matrix for galaxy surveys observables. The observables considered are the linear multitracer 3D galaxy power spectrum, the linear convergence power spectrum for weak lensing, and the linear multitracer power spectrum for the correlation between galaxy distribution and convergence. The code allows for tomographic and model-independent analysis in which, for scale-independent growth, the following functions of redshift $A_a (z) \equiv \sigma_{8}(z) \, b_a (z)$, $R(z) \equiv \sigma_{8}(z) \, f(z)$, $L(z) \equiv \Omega_{m} \, \sigma_{8}(z) \, \Sigma (z)$ and $E(z) \equiv H(z)/H_0$, together with the function of scale $\hat{P}(k)$, are taken as free parameters in each redshift and scale bins respectively. In addition, a module for change of variables is provided to project the Fisher matrix on any particular set of parameters required. The code is built to be as fast as possible and user-friendly. As an application example, we forecast the sensitivity of future galaxy surveys like DESI, Euclid, J-PAS and LSST and compare their performance on different redshift and scale ranges.

\end{abstract} 

\maketitle

\section{Introduction}

In the coming years, large-scale galaxy and lensing surveys will become one of the most powerful observational tools in cosmology. These surveys will map the 3D positions and shapes of millions of different types of galaxies and QSOs over cosmological volumes, thus allowing to measure different cosmological observables on different redshift and length scales.
Galaxy surveys can be categorized in three types according to the method used for the determination of redshifts. Thus we have spectroscopic surveys that obtain high-precision redshifts from high-quality spectra of a selected sample of galaxies. Examples of future spectroscopic surveys include the spectroscopic Euclid survey \cite{Laureijs:2011gra} and DESI \cite{Aghamousa:2016zmz}. Second, we have photometric surveys that obtain photo-spectra using photometry with a reduced number of filters. These surveys can obtain larger catalogs of sources, but at the expense of poorer redshift accuracies. Among the future photometric surveys we have the photometric Euclid survey \cite{Laureijs:2011gra} and LSST \cite{Mandelbaum:2018ouv}. Finally, we have the so-called spectro-photometric surveys that combine photometry with multi-color information obtained through the combination of broad, medium and narrow band filters. From those pseudospectra high-quality photometric redshifts can be obtaiend for a high number of sources. Examples of spectro-photometric surveys are PAU \cite{Marti:2014nha} J-PLUS \cite{Cenarro:2018uoy} and the future J-PAS \cite{Benitez:2014ibt,Bonoli:2020ciz}. 

In addition to the galaxy distribution power spectra, photometric and spectro-photometric surveys are able to measure also galaxy shapes and compute the shear or convergence power spectrum for weak lensing. Such measurements will, in particular, open the possibility to explore the properties of the dark sector
with unprecedented precision. Thus, galaxy surveys will be able to measure the equation of state of dark energy \cite{Blake:2003rh}, explore possible interactions within the dark sector \cite{Salvatelli:2014zta, Costa:2019uvk} or with the visible sector \cite{Jimenez:2020ysu} or to measure absolute neutrino masses \cite{Chudaykin:2019ock}. 
In addition, they will provide a new avenue to test gravity on cosmological scales \cite{Guzik:2009cm,Amendola:2013qna,Resco:2019xve}. In this sense it is becoming more and more useful to have tools that allow to forecast the precision with which those surveys will be able to measure certain cosmological parameters. Such forecast analysis are also very helpful in order to set the observation strategy of the survey by identifying the optimal configuration for the selected targets. With this purpose it is
useful to develop codes that are able to perform fast estimates for many different possible configurations. In this sense, thanks to its simplicity, the implementation of the Fisher matrix method can be considered as the most suitable approach for such a rapid evaluation. Fisher matrix formalism \cite{Fisher:1935cm, Tegmark:1996bz, Seo:2007ns, Heavens:2007ka, White:2008jy} assumes a Gaussian likelihood in parameter space around a fiducial model.  This approximation allows for a linear change of variable from the inverse of the data covariance matrix to the Fisher matrix for the parameters. It can be proved that this approach gives a correct order of magnitude estimate of the sizes of the parameters confidence regions  \cite{Wolz:2012sr}. 

Several Fisher matrix codes ($\texttt{BFF}$, $\texttt{CarFisher}$, $\texttt{FisherMathica}$, $\texttt{fishMath}$, $\texttt{SOAPFish}$, $\texttt{SpecSAF}$, ...) for galaxy clustering, cosmic shear and the cross correlation power spectra have been developed in the last years, although most of them are not public \cite{Blanchard:2019oqi}. These codes compute the Fisher matrix for the 3D galaxy power spectrum, the 2D galaxy power spectrum, convergence power spectrum and the cross correlation. The main approach in these codes is to numerically calculate the derivatives of the power spectra with  respect to a given set of cosmological parameters using the outputs of a Boltzmann code. 

The aim of this work is to present the new Fisher gAlaxy suRvey cOde ($\texttt{FARO}$) which is a totally public code designed to be fast and easy to use and modify. The main features of $\texttt{FARO}$ are the following:
\begin{itemize}
\item \textbf{Python code}: The code is written in Python for an easy use and manipulation. It makes extensive use of the powerful function $\texttt{np.einsum}$ and it allows to  use the Python CLASS \cite{Lesgourgues:2011re} functions to obtain the matter power spectrum.
\item \textbf{Observables}: 3D galaxy power spectrum, convergence power spectrum and cross-correlation power spectrum in the linear regime. 
\item \textbf{Multitracer}: an arbitrary number of different galaxy tracers can be considered. The corresponding multitracer galaxy-lensing cross-correlation is implemented.  
\item \textbf{Redshift and scale binning}: arbitrary number and sizes of redshift and $k$ bins can be 
chosen in an easy way.
\item \textbf{Model independent}: A set of model independent parameters are considered which allows to extract information on the constraining power of a given survey for a wide range of cosmologies. The chosen parameters
allow, in addition, to obtain the derivatives involved in the Fisher matrix calculation in an  analytical way thus making the code faster.
\item \textbf{Tomographic errors:} Error information is provided for each redhsift and $k$ bin.

\item \textbf{Flexible and user friendly}: $\texttt{FARO}$ has a simple use mode in which numerical and graphical results can be generated in a simple way. In addition, the code structure is built to be flexible and easy to modify.
\end{itemize}

The main assumptions of the code are:
\begin{itemize}
    \item \textbf{Flat FRW background}: This approximation simplifies the code and makes the calculations faster. 
    \item \textbf{Scale-independent growth factor}:
    In order to factorize the redshift and scale dependencies of the observables and 
    keep the  analysis as model-independent as possible, the growth function is assumed to be scale independent.  
\end{itemize}

Once we have described the main characteristics, we summarize here the model-independent parameters that $\texttt{FARO}$ considers. Unlike other codes which focus on particular sets of cosmological parameters for specific cosmological models such as $\Lambda$CDM, $w$CDM, modified gravity models described by a generic growth index $\gamma$, etc, $\texttt{FARO}$ uses a
set of model-independent parameters more closely related to the observables \cite{Amendola:2012ky, Amendola:2013qna}.
Thus, first of all, we introduce the redshift-dependent functions, given by
\begin{equation}\label{1.1}
A_a(z)=\sigma_{8}(z) \, b_a (z),
\end{equation}
\begin{equation}\label{1.2}
R(z)=\sigma_{8}(z) \, f(z),
\end{equation}
\begin{equation}\label{1.3}
L(z)=\Omega_{m} \, \sigma_{8}(z) \, \Sigma (z),
\end{equation}
\begin{equation}\label{1.4}
E(z)= \frac{H(z)}{H_0},
\end{equation}
where $b_a (z)$ is the bias for tracer $a$, $H(z)$ is the Hubble parameter, $\sigma_{8}(z) = \sigma_{8} \, D(z)$ being $D(z) = \delta_m (z) / \delta_m (0)$ and $\sigma_{8}$ the normalization of the matter power spectrum on scales of $8 \, \mathrm{h^{-1} Mpc}$ today; and $f(z)$ is the growth function defined as,
\begin{equation}\label{1.5}
D(z)=\exp\left[-\int_{0}^{z} \frac{f(z')}{1+z'} \, dz'\right].
\end{equation}
Finally, $\Sigma (z)$ is a general function of redshift which takes into account possible modifications of the lensing potential \cite{Pogosian:2010tj,Silvestri:2013ne} and that in $\mathrm{\Lambda CDM}$ model is $\Sigma (z) = 1$. 

Notice that we will not consider an arbitrary non-Gaussian shot noise term $P_s (z)$ as additional parameters in each redshift bin, the reasons are, on one hand,  that we are not interested in constraining them, and on the other, that it can be proved that they poorly correlated with other parameters.

In addition to the redshift-dependent functions, we have the parameters associated to the shape of the matter power spectrum. Thus we define,

\begin{equation}\label{s1.18}
\hat{P}(k)= \frac{P(k)}{\sigma_8^2},
\end{equation}
$P(k)$ being the matter power spectrum today. Notice that we have to consider $\hat{P}(k)$ as independent function instead of $P(k)$ because $\sigma_8$ is already taken into account in the redshift dependent functions (\ref{1.1}-\ref{1.4}).

Thus $\texttt{FARO}$ considers the following set of  independent parameters:  
\begin{itemize}
    \item{\it Redshift-dependent parameters}: $[A_a(z_i)$, $R(z_i)$, $L(z_i)$, $E(z_i)]$, 
where the $i$ index denotes the different redshift bins and the $a$ index the different tracers.
\item{\it Power-spectrum parameters}: $a_n$ corresponding to the amplitude of $\hat P(k)$  in the $n$-th log-spaced $k$ bin. 
\end{itemize}
Notice that this set of parameters exhausts the information that can be extracted from a galaxy and lensing survey at the linear level within the mentioned assumptions\footnote{In principle it would be possible to consider the angular diameter distances $D_A(z_i)$ as independent set of parameters from $E(z_i)$, however in our case, since the background metric is assumed to be flat FRW, this is no longer the case.}.  

There are two main aspects  that differentiate  \texttt{FARO} from other Fisher codes: on one hand the possibility of performing multitracer analysis, not only at the clustering level, but also with lensing cross-correlations, and on the other, the fact that the parametrization of the matter power-spectrum is fully model-independent. This allows \texttt{FARO} to perform forecast analysis of features and other scale-dependent deviations in the standard power-law primordial curvature spectrum or transfer function.     

The paper is organized as follows: in the first sections we explain in detail the mathematical recipes needed to compute the Fisher matrices. Thus, in \ref{secobs}, we summarize the observable power spectra that are computed, then in \ref{sec_Fisher} we explain how Fisher matrices are calculated for each observable. Finally in \ref{change} we explain how to make a change of variable from the initial parameters to a set of new  ones. We also give the example of a simple change of variable that can be done using priors. In next sections \ref{howuse} and \ref{struc}, we explain how to use $\texttt{FARO}$ in a simple way and how it is built. In \ref{secexample} we apply the code for several galaxy surveys, and in section \ref{con} we briefly discuss the results and conclusions.  Finally, several appendices are included showing the $\texttt{FARO}$ fluxchart, the specifications of the different survey analyzed and the outputs of the code. 

\section{Galaxy surveys observables}\label{secobs}

In this section we summarize the main observables of the galaxy maps that $\texttt{FARO}$ computes to obtain the Fisher matrices. These observables are the 3D galaxy power spectrum, the convergence power spectra for weak lensing and the cross-correlation power spectrum for galaxy distribution and convergence.

\subsection{Multitracer galaxy power spectrum}

One of the main observables that the code considers is the multitracer galaxy power spectrum.
We consider three effects, the linear Kaiser term for redshift space distortions \cite{Kaiser:1987qv}, the convolution redshift error term \cite{Amendola:2015ksp} and the Alcock-Paczynski effect \cite{Alcock:1979mp}. Considering these effects, the power spectrum for tracers $a$ and $b$ reads \cite{White:2008jy, McDonald:2008sh},


%
\begin{eqnarray}\label{1.6}
P_{a b}^{\delta \delta}(z,\hat{\mu}_{r},k_{r})& =& \frac{D_{A \,r}^{2} \, E}{D_A^{2} \, E_{r}} \,\, (A_a+R \, \hat{\mu}^{2}) \\ & &\times (A_b+R \, \hat{\mu}^{2}) \,\, \hat{P}(k) \,\, e^{\frac{-k_{r}^{2} \, \hat{\mu}_{r}^{2} \, \sigma_{a}^{2}}{2}} \, e^{\frac{-k_{r}^{2} \, \hat{\mu}_{r}^{2} \, \sigma_{b}^{2}}{2}}, \nonumber
\end{eqnarray}
%
where the sub-index $r$ denotes that the corresponding quantity is evaluated on the fiducial model, $\hat \mu$ is the angle of the wavevector $\vec k$ with the line of sight, $\sigma_{a}=(\delta z_a^{C} \, (1+z))/H(z)$ is the radial error for tracer $a$ where we define the error in redshift for tracer $a$ for clustering information as $\delta z_a^{C} (1+z)$. Notice that, depending on the galaxy survey, this redshift error can be a spectroscopic or a photometric error and can be different from the redshift error for lensing that we introduce below. $D_A$ is the angular distance which, in a flat Universe, reads $D_A(z)=(1+z)^{-1} \, \chi (z)$, with 
\begin{equation}\label{1.7}
\chi(z)=H_{0}^{-1} \, \int_{0}^{z} \frac{dz'}{E(z')}.
\end{equation}
The dependencies $k=k(k_{r})$, $\hat{\mu}=\hat{\mu}(\hat{\mu}_{r})$ and the factor $\frac{D_{Ar}^2 \, E}{D_A^2 \, E_{r}}$ are due to the Alcock-Paczynski effect \cite{Alcock:1979mp},
\begin{equation}\label{1.8}
k=Q \, k_{r},
\end{equation}
\begin{equation}\label{1.9}
\hat{\mu}=\frac{E \, \hat{\mu}_{r}}{E_{r} \, Q},
\end{equation}
\begin{equation}\label{1.10}
Q=\frac{\sqrt{D_{A}^{2}  \, \hat{\mu}^{2}_{r}-D_{A \,r}^{2} \, (\hat{\mu}^{2}_{r}-1)}}{D_{A \,r} }.
\end{equation}
%

\subsection{Lensing convergence power spectrum}

Galaxy maps are able to measure also the shapes and sizes of galaxies. In the standard case, this information is encoded in the convergence power spectrum (see however \cite{Resco:2018ubr,Resco:2019ena} for a more general analysis). Thus, the convergence power spectra for galaxies in redshift bins $i$ and $j$ reads \cite{Lemos:2017arq},
\begin{equation}\label{1.11}
P_{ij}^{\kappa \kappa}(\ell) = \frac{9 H_0^3}{4} \int_0^{\infty} \frac{(1+z)^2}{E(z)} g_{i} (z) g_{j} (z) L^2(z)  \hat{P}\left( \frac{\ell}{\chi(z)} \right) dz,
\end{equation}
where the window functions $g_i(z)$ are,
\begin{equation}\label{1.12}
g_{i} (z)=\int_{z}^{\infty} \left( 1-\frac{\chi(z)}{\chi(z')} \right) \, n_{i} (z') \, dz',
\end{equation}
where $n_{i} (z)$ is the galaxy density function for the $i$-bin,
\begin{equation}\label{1.13}
n_{i}(z) \propto \int_{\bar{z}_{i-1}}^{\bar{z}_{i}} n(z') \, e^{\frac{(z'-z)^{2}}{2 \sigma_{i}^{2}}} dz',
\end{equation}
being $\sigma_{i}=\delta z^{L} \, (1+ z_{i})$ the photometric redshift error for lensing information, $\bar{z}_{i}$ the upper limit of the $i$-bin, and $n(z)$ the galaxy distribution that typically reads \cite{Ma:2005rc},
\begin{equation}\label{1.14}
n(z) = \frac{3}{2 z_{p}^{3}} \, z^{2} \, e^{-(z/z_{p})^{3/2}},
\end{equation}
where $z_{p}=z_{mean}/\sqrt{2}$, being $z_{mean}$ the survey mean redshift. This is the standard galaxy distribution used for weak lensing so the code assumes it, however it is easy to modify it if necessary.

\subsection{Cross-correlation power spectra}
Apart from the 3D galaxy power spectrum defined above, it is also useful to consider the angular galaxy distribution for tracers $a$ and $b$ in redshift bins
$i$ and $j$ given by the following angular galaxy  power spectrum \cite{Guzik:2009cm, Kilbinger:2014cea},
\begin{equation}\label{1.15}
P_{ij \,\, a b}^{\delta_2 \delta_2}(\ell) = \delta_{ij} H_0 \, \frac{E(z_i)}{\chi^2(z_i)} \, A_{a} (z_i) \, A_{b} (z_j) \, \hat{P}\left( \frac{\ell}{\chi(z_i)} \right),
\end{equation}
In addition, if we have both clustering and lensing convergence information from the same volume, it is possible to
obtain the following angular cross-correlation spectrum  
\begin{equation}\label{1.16}
P_{ij \,\, a}^{\kappa \delta_2}(\ell) = \frac{3 H_0^2}{2} \, \frac{(1+z_j)}{\chi(z_j)} \, g_i (z_j) \, A_{a} (z_j) \, L(z_j) \, \hat{P}\left( \frac{\ell}{\chi(z_j)} \right),
\end{equation}
where we have assumed for the 2D galaxy distribution that galaxies in two redshift bins are not correlated with each other. 

\subsection{Matter power spectrum parametrization}

The aim of this subsection is to explain how to parametrize the normalized matter power spectrum today $\hat{P}(k)$ defined in \eqref{s1.18} in a model-independent way. We want the free parameters to be the values of $\hat{P}(k)$ in $p$ logarithmically spaced $k$ bins. However, because of the fixed normalization we will have only $p-1$ degrees of freedom to parametrize $\hat P(k)$. Considering that the fiducial model is $\mathrm{\Lambda CDM}$, a general and model-independent parametrization of $\hat{P}(k)$ with $p$ degrees of freedom can be written as,

\begin{equation}\label{s2.1}
\hat{P}(k) = g(k) \, \hat{P}_{\Lambda}(k)
\end{equation}
where $\hat{P}_{\Lambda}(k)$ is the normalized matter power spectrum for the fiducial $\mathrm{\Lambda CDM}$ model, and $g(k)$ is an arbitrary dimensionless function with the form,

\begin{equation}\label{s2.2}
g(k) = 1 + \sum_{n=0}^{p-1} \, a_n \, g_n (k).
\end{equation}

Notice that, although $g_n (k)$ with $n = 0,...,p-1$ can be general base functions, we will consider them as window functions for each logarithmic spaced $k$ bins. Due to the $\sigma_8$ constraint, these $g_n (k)$ functions cannot be independent so that the following condition is satisfied
\begin{eqnarray}\label{s2.3}
\int{\frac{k'^{2}\, dk'}{2\pi^{2}} \hat{P}(k') |\hat{W}(8,k')|^{2}} = 1,
\end{eqnarray}
where
\begin{equation}\label{s2.4}
\hat{W}(R,k)=\frac{3}{k^{3}R^{3}} \, [\sin(kR)-kR\cos(kR)]
\end{equation}
being $R = 8 \, \mathrm{Mpc/h}$. Substituting  (\ref{s2.2}) in (\ref{s2.3}) we obtain,

\begin{equation}\label{s2.5}
\sum_{n=0}^{p-1} \, a_n \, \alpha_n = 0,
\end{equation}

where,

\begin{equation}\label{s2.6}
\alpha_n \equiv \int{\frac{k'^{2}\, dk'}{2\pi^{2}} g_n (k') \, \hat{P}_{\Lambda}(k') |\hat{W}(8,k')|^{2}}.
\end{equation}

Using condition (\ref{s2.5}) we can reduce the independent parameters to $p-1$ and, without loss of generality, we can rewrite (\ref{s2.2}) as,

\begin{equation}\label{s2.7}
g(k) = 1 + \sum_{n=1}^{p-1} \, a_n \, \left[ g_n (k) - \frac{\alpha_n}{\alpha_0} \, g_0 (k) \right].
\end{equation}

Once we obtain the Fisher matrix for $a_n$ with $n = 1,...,p-1$ the corresponding errors for $\hat{P}(k)$ can be calculated by projecting the covariance matrix $C_{nm}$ using the following expression,
\begin{equation}\label{s2.8}
\frac{\sigma_{\hat{P}(k)}}{\hat{P}(k)} = \sqrt{ \left( g_n (k) - \frac{\alpha_n}{\alpha_0} \, g_0 (k) \right)  C_{nm}  \left( g_m (k) - \frac{\alpha_m}{\alpha_0} \, g_0 (k) \right)}.
\end{equation}

In particular we will consider $g_n (k)$ as a logarithmically-spaced step function of the form,

\begin{widetext}

\begin{equation}\label{s2.9}
g_n(k) = \tilde{\theta} \left( \log(k) - \left[ \log k_{n} - \frac{\Delta \log k_{n}}{2} \right] \right) \, \tilde{\theta} \left( \left[ \log k_{n} + \frac{\Delta \log k_{n}}{2} \right] - \log(k) \right),
\end{equation}

\end{widetext}
$\log k_{n}$ being the centers of the $\log(k)$ bins with size $\Delta \log k_{n}$, and $\tilde{\theta} (x)$ the Heaviside function with $\tilde{\theta} (0) = 1/2$. Using these functions , parameters $\alpha_n$ are,

\begin{equation}\label{s2.10}
\alpha_n = \int_{\log k_{n} - \frac{\Delta \log k_{n}}{2}}^{\log k_{n} + \frac{\Delta \log k_{n}}{2}} \, \frac{k'^{3}\, d\log k'}{2\pi^{2}} \, \hat{P}_{\Lambda}(k') |\hat{W}(8,k')|^{2}.
\end{equation}
%

\section{Fisher matrices}\label{sec_Fisher}

Once we have defined the observables, we will  obtain the Fisher matrix for the 3D multitracer power spectrum, the lensing convergence power spectrum and the cross-correlation power spectrum. Then, we project this Fisher matrices for the observable power spectra into the parameters we want to constrain. For the clustering multitracer power spectrum these parameters are: $[a_n, A^{a}_i, R_i, E_i]$, and for the convergence power spectra: $[a_n, E_i, L_i]$; where $a$ denotes different galaxy tracers, $i$ different redshift bins and $n$ the different $p-1$ scale bins. Finally, when we combine clustering and lensing information with the cross-correlation, the parameters are: $[a_n, A^{a}_i, R_i, L_i, E_i]$.

\subsection{Multitracer galaxy distribution Fisher matrix}\label{sec_fis_clust}

The Fisher matrix for the multitracer power spectrum reads \cite{Abramo:2011ph, Abramo:2015iga, Abramo:2019ejj},

\begin{eqnarray}\label{2.1}
&&F_{\alpha \beta}^{\delta \delta} = \sum_{i, c, m} \frac{V_i \, \Delta\hat{\mu}_m \, \Delta\log k_c \, k_c^{3}}{8 \pi^{2}} \, \left.\frac{\partial P_{a b}^{\delta \delta}(z_i,\hat{\mu}_m,k_c)}{\partial p_\alpha}\right|_{r}  \,  \\ & & \times C^{-1}_{b a'} \, \left.\frac{\partial P_{a' b'}^{\delta \delta}(z_i,\hat{\mu}_m,k_c)}{\partial p_\beta}\right|_{r}  \, C^{-1}_{b' a} \,\,\, \mathrm{e}^{-k_c^2 \, \Sigma^{2}_{\perp}-k_c^2 \, \hat{\mu}_m^2 \, (\Sigma^2_{\parallel}-\Sigma^2_{\perp})}, \nonumber
\end{eqnarray}
where we have discretized the integrals in $\hat{\mu}$ from $-1$ to $1$, and $k$ from $k_{min}$ to $\infty$. The code fixes the value $k_{min} = 0.007 \,\, h/\mathrm{Mpc}$ \cite{Amendola:2013qna}. The exponential cutoff removes the contribution from   non-linear scales \cite{Seo:2007ns}, being,
\begin{equation}\label{2.2}
\Sigma_{\perp}(z) = 0.785 \, D(z) \, \Sigma_0,
\end{equation}
\begin{equation}\label{2.3}
\Sigma_{\parallel}(z) = 0.785 \, D(z) \, (1+f(z)) \, \Sigma_0.
\end{equation}
The matrix $C_{a b}$ is
\begin{equation}\label{2.4}
C_{a b}(z_i,\hat{\mu}_m,k_c) = P_{a b}^{\delta \delta}(z_i,\hat{\mu}_m,k_c) + \frac{\delta_{a b}}{\bar{n}_a(z_i)},
\end{equation}
where $\bar{n}_a (z_i)$ is the galaxy density of tracer $a$ in the $z$ bin $i$. Finally, $V_{i}$ is the total volume of the $\textrm{i}$-th bin. For a flat $\mathrm{\Lambda CDM}$ model,  $V_{i}=\frac{4 \pi \, f_{sky}}{3} \, \left(\chi(\bar{z}_{i})^{3}-\chi(\bar{z}_{i-1})^{3}\right)$ where $f_{sky}$ is the sky fraction of the survey and $\bar{z}_{i}$ the upper limit of the $i$-th bin.

Now we show the derivatives with respect to the parameters $[a_n, A_{k}^l, R_{k}, E_{k}]$, where $k$ denotes different redshift bins, $l$ different galaxy tracers and $n$ the $p-1$ different scale bins,
\begin{widetext}
\begin{eqnarray}\label{2.5.1}
\left.\frac{\partial P_{a b}^{\delta \delta}(z_i,\hat{\mu}_m,k_c)}{\partial a_n}\right|_{r} = \left[ g_n (k_c) - \frac{\alpha_n}{\alpha_0} \, g_0 (k_c) \right] \, P_{a b}^{\delta \delta}(z_i,\hat{\mu}_m,k_c),
\end{eqnarray}
\begin{eqnarray}\label{2.5}
\left.\frac{\partial P_{a b}^{\delta \delta}(z_i,\hat{\mu}_m,k_c)}{\partial A_k^l}\right|_{r} = \left[ \frac{\delta_{l a} \delta_{k i}}{A_i^a + R_i \, \hat{\mu}_m^2} + \frac{\delta_{l b} \delta_{k i}}{A_i^b + R_i \, \hat{\mu}_m^2} \right] \, P_{a b}^{\delta \delta}(z_i,\hat{\mu}_m,k_c),
\end{eqnarray}
\begin{eqnarray}\label{2.6}
\left.\frac{\partial P_{a b}^{\delta \delta}(z_i,\hat{\mu}_m,k_c)}{\partial R_k}\right|_{r} = \left[ \frac{\delta_{k i} \, \hat{\mu}_m^2}{A_i^a + R_i \, \hat{\mu}_m^2} + \frac{\delta_{k i} \, \hat{\mu}_m^2}{A_i^b + R_i \, \hat{\mu}_m^2} \right] \, P_{a b}^{\delta \delta}(z_i,\hat{\mu}_m,k_c),
\end{eqnarray}
\begin{align}\label{2.7}
\left.\frac{\partial P_{a b}^{\delta \delta}(z_i,\hat{\mu}_m,k_c)}{\partial E_k}\right|_{r} =& \left[ \frac{\delta_{k i}}{E_k} - \frac{2}{\chi(z_i)} \frac{\partial \chi (z_i)}{\partial E_{k}} + \frac{d \log \hat P}{d \log k} (k_c) \, \left( \frac{\hat{\mu}_m^2 \delta_{k i}}{E_k} - \frac{(1-\hat{\mu}_m^2)}{\chi (z_i)} \frac{\partial \chi (z_i)}{\partial E_{k}} \right) \right. \\ \nonumber
&\left. + \left( \frac{2 R_i \hat{\mu}_m^2 (1-\hat{\mu}_m^2)}{A_i^a + R_i \, \hat{\mu}_m^2} + \frac{2 R_i \hat{\mu}_m^2 (1-\hat{\mu}_m^2)}{A_i^b + R_i \, \hat{\mu}_m^2} \right) \times \left( \frac{\delta_{k i}}{E_k} + \frac{1}{\chi(z_i)} \frac{\partial \chi (z_i)}{\partial E_{k}} \right) \right] \, P_{a b}^{\delta \delta}(z_i,\hat{\mu}_m,k_c),
\end{align}
\end{widetext}
evaluated on the fiducial model, where,
\begin{eqnarray}\label{2.8}
\frac{\partial \chi (z_i)}{\partial E_{k}} = - \theta (z_i - z_k) \, \frac{\Delta z_k}{H_0 \, E_{k}^2},
\end{eqnarray}
being $\theta (x)$ the Heaviside function with $\theta (0) = 1$. The resulting Fisher matrix has the following form for parameters $[a_n, A_{i}^a, R_{i}, E_{i}]$,
\[ \left( \begin{array}{ccccccccc}
 a_1 a_1 & a_1 a_2 & ... & a_1 A_1^1 & a_1 A_1^2 & ... & a_1 R_1 & a_1 E_1 & ...  \\
 a_2 a_1 & a_2 a_2 & ... & a_2 A_1^1 & a_2 A_1^2 & ... & a_2 R_1 & a_2 E_1 & ...  \\
 ... & ... & ... & ... & ... & ... & ... & ... & ... \\
 A_1^1 a_1 & A_1^1 a_2 & ... & A_1^1 A_1^1 & A_1^1 A_1^2 & ... & A_1^1 R_1 & A_1^1 E_1 & ...  \\
 A_1^2 a_1 & A_1^2 a_2 & ... & A_1^2 A_1^1 & A_1^2 A_1^2 & ... & A_1^2 R_1 & A_1^2 E_1 & ...  \\
 ... & ... & ... & ... & ... & ... & ... & ... & ... \\
 R_1 a_1 & R_1 a_2 & ... & R_1 A_1^1 & R_1 A_1^2 & ... & R_1 R_1 & R_1 E_1 & ...  \\
 E_1 a_1 & E_1 a_2 & ... & E_1 A_1^1 & E_1 A_1^2 & ... & E_1 R_1 & E_1 E_1 & ...  \\
 ... & ... & ... & ... & ... & ... & ... & ... & ... \\
\end{array}\; \right)\;.\]
%

\subsection{Convergence power spectrum Fisher matrix}\label{sec_fis_lens}

The Fisher matrix for the convergence power spectrum reads \cite{Eisenstein:1998hr},
\begin{eqnarray}\label{2.9}
F_{\alpha \beta}^{\kappa \kappa} = f_{sky}  \sum_{\ell} \Delta \ln \ell & & \frac{(2\ell+1)  \ell}{2} \left.\frac{\partial P_{i j}^{\kappa \kappa}(\ell)}{\partial p_\alpha}\right|_{r} \\ \nonumber & & \times C^{-1}_{j i'} \, \left.\frac{\partial P_{i' j'}^{\kappa \kappa}(\ell)}{\partial p_\beta}\right|_{r}  \, C^{-1}_{j' i},
\end{eqnarray}
with
\begin{equation}\label{2.10}
C_{ij} (\ell)=P_{ij}^{\kappa \kappa}(\ell)+\gamma_{int}^{2} \, \hat{n}_{i}^{-1} \, \delta_{ij},
\end{equation}
where $\gamma_{int}$ is the intrinsic ellipticity, and $\hat{n}_{i}$ the galaxies per steradian in the $i$-th bin,
\begin{equation}\label{2.11}
\hat{n}_{i}=n_{\theta} \, \frac{\int_{\bar{z}_{i-1}}^{\bar{z}_{i}} n(z) \, dz}{\int_{0}^{\infty} n(z) \, dz},
\end{equation}
where $n_{\theta}$ is the areal galaxy density and $n(z)$ follows (\ref{1.14}). We sum in $\ell$ with $\Delta \ln \ell=0.1$ from $\ell_{\text{min}}=5$ \cite{Amendola:2013qna} to $\ell_{\text{max}}$ with
$\ell_{\text{max}}= \chi(z_{\alpha'}) \, k_{\text{max}}$ where $\alpha' = \mathrm{min}(\alpha,\beta)$ and $k_{\text{max}}(z_{i})$ is defined so that $\sigma(z_{i},\pi/2k_{\text{max}}(z_{i}))=0.35$ being,
\begin{eqnarray}\label{2.12}
\sigma^{2}(z,R)= \sigma_8^{2}(z) \int{\frac{k'^{2}\, dk'}{2\pi^{2}} \hat{P}(k') |\hat{W}(R,k')|^{2}},
\end{eqnarray}
where we use a top-hat filter $\hat{W}(R,k)$ defined by (\ref{s2.4}). So that we are  only considering modes in the linear regime. Now we show the derivatives with respect to the parameters $[a_n, E_{k}, L_{k}]$,
\begin{widetext}
\begin{equation}\label{2.14.1}
\frac{\partial P_{ij}^{\kappa \kappa}(\ell)}{\partial a_{n}} = \sum_{i'} \, \left[ g_n \left(\frac{\ell}{\chi (z_{i'})}\right) - \frac{\alpha_n}{\alpha_0} \, g_0 \left(\frac{\ell}{\chi (z_{i'})}\right) \right] \, P_{ij}^{\kappa \kappa}(z_{i'}, \ell),
\end{equation}
\begin{eqnarray}\label{2.14}
\frac{\partial P_{ij}^{\kappa \kappa}(\ell)}{\partial E_{k}}&=& -\frac{1}{E_k} \, P_{ij}^{\kappa \kappa}(z_k, \ell) 
+ \sum_{i'} \, \frac{1}{g_i (z_{i'})} \, \frac{\partial g_i (z_{i'})}{\partial E_k} \, P_{ij}^{\kappa \kappa}(z_{i'}, \ell) 
+ \sum_{i'} \, \frac{1}{g_j (z_{i'})} \, \frac{\partial g_j (z_{i'})}{\partial E_k} \, P_{ij}^{\kappa \kappa}(z_{i'}, \ell) \\ \nonumber
&-& \sum_{i'} \, \frac{1}{\chi (z_{i'})} \, \frac{\partial \chi (z_{i'})}{\partial E_k} \, \frac{d \log \hat P}{d \log k} \left(\frac{\ell}{\chi (z_{i'})}\right) \, P_{ij}^{\kappa \kappa}(z_{i'}, \ell),
\end{eqnarray}
\begin{equation}\label{2.15}
\frac{\partial P_{ij}^{\kappa \kappa}(\ell)}{\partial L_{k}}=\frac{2}{L_k} \, P_{ij}^{\kappa \kappa}(z_k, \ell),
\end{equation}
where,
\begin{eqnarray}\label{2.16}
P_{ij}^{\kappa \kappa}(z_k, \ell) = \frac{9 H_0^3}{4} \frac{(1+z_k)^2 \Delta z_k}{E_k} g_{i} (z_k) g_{j} (z_k) \, L_k^{2} \, \hat{P}\left( \frac{\ell}{\chi(z_k)} \right),
\end{eqnarray}
and
\begin{eqnarray}\label{2.17}
\frac{\partial g_{i} (z_{j})}{\partial E_{k}}=\frac{\Delta z_{k}}{H_0 \, E_{k}^{2}} \left[ - \hat{\theta} (z_{k}-z_{j}) \chi (z_{j})  \int_{z_{k}}^{\infty} \frac{n_{i} (z')}{\chi (z')^{2}} \, dz'  
 + \theta (z_{j}-z_{k}) \int_{z_{j}}^{\infty} \left(1-\frac{\chi (z_{j})}{\chi (z')} \right) \frac{n_{i} (z')}{\chi (z')}dz' \right],
\end{eqnarray}
\end{widetext}
with $\hat{\theta}(0)=0$, $\theta(0)=1$ and $\Delta z_{k}$ the size of the redshift bin $z_{k}$. The resulting Fisher matrix has the following form for parameters $[a_n, E_{i}, L_{i}]$,

\[ \left( \begin{array}{ccccccccc}
 a_1 a_1 & a_1 a_2 & ... & a_{1}E_{1} & a_{1}L_{1} & a_{1}E_{2} & a_{1}L_{2} & ... \\
 a_2 a_1 & a_2 a_2 & ... & a_{2}E_{1} & a_{2}L_{1} & a_{2}E_{2} & a_{2}L_{2} & ... \\
 ... & ... & ... & ... & ... & ... & ... & ... \\
E_1 a_1  & E_1 a_2 & ... & E_{1}E_{1} & E_{1}L_{1} & E_{1}E_{2} & E_{1}L_{2} & ... \\
L_1 a_1 & L_1 a_2 & ... & L_{1}E_{1} & L_{1}L_{1} & L_{1}E_{2} & L_{1}L_{2} & ... \\
E_2 a_1 & E_2 a_2 & ... & E_{2}E_{1} & E_{2}L_{1} & E_{2}E_{2} & E_{2}L_{2} & ... \\
L_2 a_1 & L_2 a_2 & ... & L_{2}E_{1} & L_{2}L_{1} & L_{2}E_{2} & L_{2}L_{2} & ... \\
 ... & ... & ... & ... & ... & ... & ... & ... \\
\end{array}\; \right)\;.\]
%

\subsection{Cross-correlation power spectrum Fisher matrix}

The Fisher matrix for the cross-correlation power spectrum reads \cite{Guzik:2009cm},
\begin{eqnarray}\label{2.18}
F_{\alpha \beta}^{\kappa \delta_2} = f_{sky}  \sum_{\ell} \Delta \ln \ell & & \frac{(2\ell+1)  \ell}{2} \left.\frac{\partial P_{i j \, a}^{\kappa \delta_2}(\ell)}{\partial p_\alpha}\right|_{r} \\ \nonumber & & \times \left[ C_{iji'j'}^{a b} \right]^{-1} \, \left.\frac{\partial P_{i' j' \, b}^{\kappa \delta_2}(\ell)}{\partial p_\beta}\right|_{r},
\end{eqnarray}
with,
\begin{equation}\label{2.19}
C_{iji'j'}^{a b} (\ell) = C_{i i'} (\ell) C_{j j'}^{a b} (\ell) + C_{i j'}^{a} (\ell) C_{j i'}^{b} (\ell),
\end{equation}
where $C_{i i'} (\ell)$ follows (\ref{2.10}), $C_{i j'}^{a} (\ell) = P_{i j' \,\, a}^{\kappa \delta_2}(\ell)$ and,
\begin{equation}\label{2.20}
C_{j j'}^{a b} (\ell) = P_{j j' \,\, a b}^{\delta_2 \delta_2}(\ell) + \frac{\delta_{j j'} \delta_{a b}}{\hat{n}_i^{a}},
\end{equation}
being $\hat{n}_i^{a}$ the galaxies per steradian in the $i$-th bin for the tracer $a$. We calculate this number of galaxies using the galaxy density $\bar{n}_a (z_i)$,
\begin{equation}\label{2.21}
\hat{n}_i^{a} = \frac{1}{3} \, \left(\chi(\bar{z}_{i})^{3}-\chi(\bar{z}_{i-1})^{3}\right) \, \bar{n}_a (z_i).
\end{equation}
As we can see in (\ref{1.16}), the cross-correlation power spectrum depends on the product $Q_{a} (z) \equiv A_{a} (z) \, L(z)$. It can be proved that the corresponding Fisher matrix involves only two independent parameters in each redshift bin: $[Q_{a} (z_i), E(z_i)]$. The degeneracy of $A_{a} (z)$ and $L(z)$ can be broken when we combine the Fisher matrices for clustering and lensing convergence. Since we are not interested in the information of the cross-correlation power spectrum alone, and we will combine with other Fisher matrices, we project the initial Fisher matrix of $[Q_{a} (z_i), E(z_i)]$ into a Fisher matrix of $[A_{a} (z_i), L(z_i), E(z_i)]$ and then we combine it with the Fisher matrices for clustering $[A_{a} (z_i), R(z_i), E(z_i)]$, and lensing $[E(z_i), L(z_i)]$. 

In addition to the redshift-dependent parameters, we also have the scale dependence of $\hat{P}(k)$. Thus, the derivatives of  the cross-correlation power spectrum with respect to the parameters $[a_n, A_{k}^l, L_{k}, E_{k}]$, where $k$ denotes different redshift bins, $l$ different galaxy tracers and $n$ the $p-1$ different scale bins, are,
\begin{widetext}
\begin{equation}\label{2.22.1}
\frac{\partial P_{ij \, a}^{\kappa \delta_2}(\ell)}{\partial a_n}= \left[ g_n \left(\frac{\ell}{\chi (z_j)}\right) - \frac{\alpha_n}{\alpha_0} \, g_0 \left(\frac{\ell}{\chi (z_j)}\right) \right] \, P_{ij \, a}^{\kappa \delta_2}(\ell),
\end{equation}
\begin{equation}\label{2.22}
\frac{\partial P_{ij \, a}^{\kappa \delta_2}(\ell)}{\partial A_k^l}=\frac{\delta _{k j} \, \delta _{a l}}{A_k^l} \, P_{ij \, a}^{\kappa \delta_2}(\ell),
\end{equation}
\\
\begin{equation}\label{2.23}
\frac{\partial P_{ij \, a}^{\kappa \delta_2}(\ell)}{\partial L_{k}}=\frac{\delta _{k j}}{L_k} \, P_{ij \, a}^{\kappa \delta_2}(\ell),
\end{equation}
\begin{eqnarray}\label{2.24}
\frac{\partial P_{ij \, a}^{\kappa \delta_2}(\ell)}{\partial E_{k}}&=& \left[ \frac{1}{g_i (z_j)} \frac{\partial g_i (z_j)}{\partial E_{k}} - \frac{1}{\chi(z_j)} \frac{\partial \chi (z_j)}{\partial E_{k}} 
 - \frac{1}{\chi (z_j)} \, \frac{\partial \chi (z_j)}{\partial E_k} \, \frac{d \log \hat P}{d \log k} \left(\frac{\ell}{\chi (z_j)}\right)  \right] \, P_{ij \, a}^{\kappa \delta_2}(\ell).
\end{eqnarray}
The total Fisher matrix for clustering and weak lensing considering the cross-correlation power spectrum has the following form for parameters $[a_n, A_{i}^a, R_{i}, L_{i}, E_{i}]$,

\[ \left( \begin{array}{cccccccccc}
a_1 a_1 & a_1 a_2 & ... & a_1 A_1^1 & a_1 A_1^2 & ... & a_1 R_1 & a_1 L_1 & a_1 E_1 & ...  \\
a_2 a_1 & a_2 a_2 & ... & a_2 A_1^1 & a_2 A_1^2 & ... & a_2 R_1 & a_2 L_1 & a_2 E_1 & ...  \\
... & ... & ... & ... & ... & ... & ... & ... & ... & ... \\
A_1^1 a_1 & A_1^1 a_2 & ... & A_1^1 A_1^1 & A_1^1 A_1^2 & ... & A_1^1 R_1 & A_1^1 L_1 & A_1^1 E_1 & ...  \\
A_1^2 a_1 & A_1^2 a_2 & ... & A_1^2 A_1^1 & A_1^2 A_1^2 & ... & A_1^2 R_1 & A_1^2 L_1 & A_1^2 E_1 & ...  \\
... & ... & ... & ... & ... & ... & ... & ... & ... & ... \\
R_1 a_1 & R_1 a_2 & ... & R_1 A_1^1 & R_1 A_1^2 & ... & R_1 R_1 & 0 & R_1 E_1 & ...  \\
L_1 a_1 & L_1 a_2 & ... & L_1 A_1^1 & L_1 A_1^2 & ... & 0 & L_1 L_1 & L_1 E_1 & ...  \\
E_1 a_1 & E_1 a_2 & ... & E_1 A_1^1 & E_1 A_1^2 & ... & E_1 R_1 & E_1 L_1 & E_1 E_1 & ...  \\
... & ... & ... & ... & ... & ... & ... & ... & ... & ... \\
\end{array}\; \right)\;.\]
\end{widetext}
\section{Change of variable module} \label{change}

The change of variable module can be used to project from the initial set of  parameters to the desired ones. Given an initial Fisher matrix for parameters $\{p_\alpha\}$, we can obtain the Fisher matrix for a new set of parameters $\{q_\alpha\}$ as,
\begin{equation}\label{3.1}
\textbf{F}^{q}=\textbf{P}^{t} \, \textbf{F}^{p} \, \textbf{P},
\end{equation}
where $P_{\alpha\beta}=\partial{p_{\alpha}}/\partial{q_{\beta}}$ is evaluated on the fiducial model. The module is built as general as possible to make an arbitrary change of variables. To illustrate how to use it, the code contains an explicit example of a change of variable that we explain below.

\subsection*{Example: breaking degeneracies with $\sigma_8$ priors}\label{mod}

The redshift-dependent parameters that we are considering
$\left[\, \sigma_8(z)b_a(z), \sigma_8(z)f(z), \Omega_M\sigma_8(z)\Sigma (z), E(z)\,\right]$, 
exhibit degeneracies among different pairs of  parameters. In particular,
 $\sigma_8(z)$ and $b_a(z)$, $\sigma_8(z)$ and $f(z)$ or
 $\sigma_8(z)$ and $\Sigma(z)$. In flat  $\mathrm{\Lambda CDM}$ since the $E(z)$ measurement fixes $\Omega_M$ and $\Sigma(z)=1$,  $L(z)$ allows to determine $\sigma_8(z)$. In such a case, it is possible to break the above mentioned degeneracies and
 determine observationally $\left[\,b_a(z), f(z), \sigma_8 (z), E(z)\,\right]$. In more general cosmologies with $\Sigma(z)\neq 1$, as is the case of modified gravity models, this is no longer possible and we need additional information to break the mentioned degeneracies. We can consider, for example,  the Planck measurement of the matter power spectrum amplitude, $\sigma_8$, as a prior. Introducing this prior in the Fisher matrix and using the relation (\ref{1.5}), we can obtain the Fisher matrix for  $\left[\,b_a(z), f(z), \Sigma (z), E(z)\,\right]$.
 For that purpose, we need the following non-zero derivatives to perform the change of variable from the initial variables (\ref{1.1}-\ref{1.4}) to the new ones, 
\begin{equation}\label{3.2}
\frac{\partial A_i^{a}}{\partial \sigma_8} = \frac{1}{\sigma_8} \, A_i^{a},
\end{equation}
\begin{equation}\label{3.3}
\frac{\partial R_i}{\partial \sigma_8} = \frac{1}{\sigma_8} \, R_i,
\end{equation}
\begin{equation}\label{3.4}
\frac{\partial L_i}{\partial \sigma_8} = \frac{1}{\sigma_8} \, L_i,
\end{equation}
\begin{equation}\label{3.5}
\frac{\partial A_i^{a}}{\partial b_j^{b}} = \frac{\delta_{ij} \, \delta_{a b}}{b_j^{b}} \, A_i^{a},
\end{equation}
\begin{equation}\label{3.6}
\frac{\partial A_i^{a}}{\partial f_j} = \frac{A_i^{a}}{D_i} \, \frac{\partial D_i}{\partial f_j},
\end{equation}
\begin{equation}\label{3.7}
\frac{\partial R_i}{\partial f_j} = \sigma_8 \, D_i \, \delta_{i j} + \sigma_8 \, f_i \, \frac{\partial D_i}{\partial f_j},
\end{equation}
\begin{equation}\label{3.8}
\frac{\partial L_i}{\partial f_j} = \frac{L_i}{D_i} \, \frac{\partial D_i}{\partial f_j},
\end{equation}
\begin{equation}\label{3.9}
\frac{\partial L_i}{\partial \Sigma_j} = L_i \, \delta_{i j},
\end{equation}

being,

\begin{equation}\label{3.10}
\frac{\partial D_i}{\partial f_j} =
    \begin{cases}
      - \frac{D_i \, \Delta z_j}{1+z_j}, & \text{if}\  i \geq j \\
      0, & \text{if}\ j > i
    \end{cases}
\end{equation}

This explicit change of variable has been  implemented explicitly in the code as an example.

\section{How to use $\texttt{FARO}$ in a simple way}\label{howuse}

The last sections are dedicated to explain how the code works from the simplest to the more complete way. First of all, we explain how to use it and what are the generated output files. In order to use \texttt{FARO} the \texttt{"Input\_file.py"} module has to be completed with the required inputs and run \texttt{"FARO.py"} in the same folder. Let us summarize the required inputs of the $\texttt{"Input\_file.py"}$ module:   

\begin{itemize}
\item \textbf{Name of the survey}. A text variable with the survey name. The code name is \texttt{name}.
\item \textbf{Observable used}. This variable set the power spectra we want to use to compute the Fisher matrix. There are three possible choices: to compute the multitracer power spectrum (code name $\texttt{do = 'C'}$), the lensing convergence power spectrum (code name \texttt{do = 'L'}) or the combination of clustering and lensing with the cross-correlated power spectrum (code name \texttt{do = 'C\_L\_C'}).
\item \textbf{Fiducial cosmology}. Values for the fiducial cosmology parameters: $(\Omega_m, \gamma, h, \omega_b, n_s, \sigma_8)$ with code names: \texttt{(Omegam, gamma, h, omegab, ns, s8)}.
\item \textbf{Redshift bins and fraction of the sky:}
\begin{itemize}
\item \textbf{Redshift centers}: a numpy array with the centers of redshift bins, the code name is $\texttt{z}$. For example: "\texttt{z = np.array([0.3, 0.5, 0.9])}".
\item \textbf{Lower and upper limits of redshift bins}: a numpy array with the lower and upper limits of each redshift bin, the code name is \texttt{zb}. For example: "\texttt{zb = np.array([0.2, 0.4, 0.6, 1.2])}". Notice that, although redshift centers and the edges are not independent, we include both explicitly to avoid confusions.
\item \textbf{Fraction of the sky}: the code name is \texttt{fsky}. You can also use as input the number of square degrees of the survey, in that case $f_{sky} = \left(2.42 \times 10^{-5}\right) \, \mathrm{deg^2}$.
\end{itemize}
\item \textbf{Log-spaced $k$-bins:}
\begin{itemize}
\item \textbf{$\log k_a$ centers}: a numpy array with the centers of $\log k$ bins, the code name is \texttt{lnk}.
\item \textbf{Lower and upper limits of $\log k$ bins}: a numpy array with the lower and upper limits of each $\log k$ bin, the code name is $\texttt{lnkb}$. By default, the code takes $10$ bins from $k = 0.007 \, \mathrm{h/Mpc}$ to $k = 1 \, \mathrm{h/Mpc}$. Notice that, although $\log k_a$ centers and limits are not independent, we include both explicitly to avoid any confusion.
\end{itemize}
\item \textbf{Clustering features}. These variables are only used when \texttt{do = 'C'} or \texttt{do = 'C\_L\_C'}.
\begin{itemize}
\item \textbf{Non-linear cutoff}. Value for $\Sigma_0$ in units of $\mathrm{Mpc/h}$. The code name is \texttt{S0}.
\item \textbf{Galaxy densities}: a numpy matrix with one row (galaxy density in each redshift bin in  $(h/\text{Mpc})^3$ units) for each tracer. The code name is $\texttt{nb}$. For example: "\texttt{nb = np.array([n\_tra1, n\_tra2, ...])}".
\item \textbf{Redshift errors for clustering}: a numpy array with the clustering redshift error for each tracer, the code name is $\texttt{dzC}$. For example: "\texttt{dzC = np.array([dz\_tra1, dz\_tra2, ..])}".
\item \textbf{Fiducial bias}: a numpy matrix with one row (bias in each redshift bin) for each tracer, the code name is \texttt{b}. For example: \texttt{b = np.array([b\_tra1, b\_tra2, ...])}. The bias values in each redshift bin are computed using the function \texttt{b\_z( z, Omegam, gamma )} from the \texttt{Functions.py} module. A new bias function can be added to this function if  needed. 
\item \textbf{Bias names}: a numpy array with the names of each galaxy tracer. The code name is \texttt{bias\_names}. For example: "\texttt{bias\_names = np.array(['name\_tra1', ...])}".
\end{itemize}
\item \textbf{Lensing features}. These variables are only used when $\texttt{do = 'L'}$ or \texttt{do = 'C\_L\_C'}.
\begin{itemize}
\item \textbf{Intrinsic ellipticity}: value of $\gamma_{int}$. The code name is $\texttt{gint}$.
\item \textbf{Redshift error for lensing}: value for the lensing redshift error. The code name is \texttt{dzL}.
\item \textbf{Areal galaxy density}: value of the areal galaxy density $n_{\theta}$ in units of galaxies per square arc minute. The code name is \texttt{nt}.
\item \textbf{Survey mean redshift}: value of the survey mean redshift $z_{mean}$. The code name is $\texttt{zm}$.
\end{itemize}
\end{itemize}

Once we have entered our inputs in \texttt{"Input\_file.py"} and  \texttt{"FARO.py"} is run, the outputs are generated in the folder \texttt{.\slash Results}. If we have selected 
\texttt{do = 'C'} the results are in \texttt{.\slash Results\slash Clustering\_only\slash}, if \texttt{do = 'L'} they are in \texttt{.\slash Results\slash Weak\_lensing\_only\slash} and finally if \texttt{do = 'C\_L\_C'}  they are in \texttt{.\slash Results\slash Clustering\_lensing\_cross\slash}. Now we summarize the outputs in each case.

\subsection{Multitracer power spectrum outputs}

A folder with the name of the survey is created, with two additional sub-folders: \texttt{Model\_independent} and \texttt{Change\_variable}; with four additional folders inside each one: \texttt{Data\_survey}, \texttt{Fisher\_matrices}, \texttt{Plots} and \texttt{Tables}.
\begin{itemize}
\item \textbf{Data\_survey}: here a file named \texttt{"Survey name"\_data.npz} is generated. This file can be read with the \texttt{Data\_import.py} module to recover the survey inputs.
\item \textbf{Fisher\_matrices}: in \texttt{Model\_independent\slash Fisher\_matrices} a file is created named \texttt{"Survey name"\_Fisher\_p\_A\_R\_E.out} with the Fisher matrices of the multitracer power spectrum for parameters $[a_n, A_{i}^a, R_{i}, E_{i}]$. In \texttt{Change\_variable\slash Fisher\_matrices} a file is created named \texttt{"Survey name"\_Fisher\_b\_f\_E.out} with the Fisher matrices of the multitracer power spectrum for parameters $[b_{i}^a, f_{i}, E_{i}]$. The Fisher matrix of $[b_{i}^a, f_{i}, E_{i}]$ has been marginalized with  respect to the $a_n$ parameters.
\item \textbf{Plots}: graphs are generated in \texttt{.pdf} format for the percentage relative errors of parameters $A_a(z)$, $R(z)$, $E(z)$ and $\hat{P}(k)$ in \texttt{Model\_independent\slash Plots}; and for $b_a(z)$ and $f(z)$ in \texttt{Change\_variable\slash Plots}.
\item \textbf{Tables}: In \texttt{.\slash Model\_independent\slash Tables} two files named \texttt{"Survey name"\_errors\_A\_R\_E.out} and \texttt{"Survey name"\_errors\_P\_k.out} are created with the values of redshift bins, fiducial values, absolute and relative errors of $[A_{i}^a, R_{i}, E_{i}]$ and with the values of $k$, fiducial values, absolute and relative errors of $\hat{P}(k)$ in units of $h/\mathrm{Mpc}$. In \texttt{.\slash Change\_variable\slash Tables} one file named \texttt{"Survey name"\_errors\_b\_f\_E.out} is created with the values of redshift bins, fiducial values, absolute and relative errors of $[b_{i}^a, f_{i}, E_{i}]$.
\end{itemize}

\subsection{Lensing convergence power spectrum outputs}

A folder with the name of the survey is created again with four additional folders inside: \texttt{Data\_survey}, \texttt{Fisher\_matrices}, \texttt{Plots} and \texttt{Tables}.
\begin{itemize}
\item \textbf{Data\_survey}: here a file named \texttt{"Survey name"\_data.npz} is generated. This file can be read with the \texttt{Data\_import.py} module to recover the survey inputs.
\item \textbf{Fisher\_matrices}: one file named \texttt{"Survey name"\_Fisher\_p\_E\_L.out} is created with the Fisher matrix of the convergence power spectrum for parameters $[a_n, E_{i}, L_{i}]$ following the structure of \ref{sec_fis_lens}.
\item \textbf{Plots}: graphs are generated in \texttt{.pdf} format for the percentage relative errors of parameters $E(z)$, $L(z)$ and $\hat{P}(k)$.
\item \textbf{Tables}: two files named \texttt{"Survey name"\_errors.out} and \texttt{"Survey name"\_errors\_P\_k.out} are created with the values of redshift bins, fiducial values, absolute and relative errors of $L_{i}$ and $E_{i}$ in the first file, and with the values of $k$, fiducial values, absolute and relative errors of $\hat{P}(k)$ in units of $h/\mathrm{Mpc}$ in the last file.
\end{itemize}

\subsection{Clustering, lensing and cross-correlation power spectra outputs}

A folder with the name of the survey is created, with two additional sub-folders: \texttt{Model\_independent} and \texttt{Change\_variable}; with four additional folders inside each one: \texttt{Data\_survey}, \texttt{Fisher\_matrices}, \texttt{Plots} and \texttt{Tables}.
\begin{itemize}
\item \textbf{Data\_survey}: here a file named \texttt{"Survey name"\_data.npz} is generated. This file can be read with the \texttt{Data\_import.py} module to recover the survey inputs.
\item \textbf{Fisher\_matrices}: In \texttt{Model\_independent\slash Fisher\_matrices} three files named \texttt{"Survey name"\_Fisher\_clust\_p\_A\_R\_E.out}, \texttt{"Survey name"\_Fisher\_lens\_p\_E\_L.out} and \texttt{"Survey name"\_Fisher\_tot\_p\_A\_R\_L\_E.out} are created with the Fisher matrices of the clustering alone, lensing alone and the combination with the cross-correlation power spectrum. In \texttt{Change\_variable\slash Fisher\_matrices} two files named \texttt{"Survey name"\_Fisher\_clust\_b\_f\_E.out} and \texttt{"Survey name"\_Fisher\_tot\_b\_f\_Sigma\_E.out} are created with the Fisher matrices of the clustering alone, and the combination of clustering, lensing and the cross correlation for the change of variable case. The structure of these matrices are explained in detail in section (\ref{sec_Fisher}).
\item \textbf{Plots}: In \texttt{Model\_independent\slash Plots} graphs are generated in \texttt{.pdf} format for the percentage relative errors of parameters $A_a(z)$, $R(z)$, $L(z)$, $E(z)$ and $\hat{P}(k)$. In \texttt{Change\_variable\slash Plots} graphs are generated in \texttt{.pdf} format for the percentage relative errors of parameters $b_a(z)$, $f(z)$ and $\Sigma(z)$.
\item \textbf{Tables}: In \texttt{Model\_independent\slash Tables} four files named \texttt{"Survey name"\_errors\_clust\_A\_R\_E.out}, \texttt{"Survey name"\_errors\_lens.out}, \texttt{"Survey name"\_errors\_tot\_A\_R\_L\_E.out} and \texttt{"Survey name"\_errors\_P\_k.out} are created with the values of redshift bins, fiducial values, absolute and relative errors for $A_a(z)$, $R(z)$, $L(z)$ and $E(z)$;  with the clustering information, the lensing information and the total combination in the first three files. Finally, the last file contains the values of $k$, the fiducial values, the absolute and relative errors of $\hat{P}(k)$ in units of $h/\mathrm{Mpc}$ for clustering information, lensing information and the total combination. In \texttt{Change\_variable\slash Tables} two files named \texttt{"Survey name"\_errors\_clust\_b\_f\_E.out} and \texttt{"Survey name"\_errors\_tot\_b\_f\_Sigma\_E.out} are created with the values of the redshift bins, the fiducial values, the absolute and relative errors for $b_a(z)$, $f(z)$, $\Sigma(z)$ and $E(z)$ obtained from the clustering information, the lensing information and the total combination.
\end{itemize}

\section{Code structure}\label{struc}

Here we explain how \texttt{FARO} is structured and the content of each folder. The code is designed in such a way that all the basic functions are included in an independent module. We also describe the modules included in each of the four folders that constitutes the code. The structure and the relation between different modules are represented in the flux chart in Fig. \ref{Figure_1}.

\begin{itemize}
\item \texttt{.\slash FARO\slash}: in this folder we have the modules \texttt{Input\_file.py} and \texttt{FARO.py}. These modules are used to enter the inputs and run the code in a simple way.

\item \texttt{.\slash FARO\slash Modules\slash}: in this folder we have the modules \texttt{Plot\_table\_gen.py} and \texttt{Data\_Import.py}. The purpose of these modules is to manipulate the results of the Fisher matrices and generate the tables and plots of the outputs.

\item \texttt{.\slash FARO\slash Modules\slash Change\_control\slash}: in this folder we have the modules \texttt{Control.py}, \texttt{Change.py} and \texttt{Change\_b\_f\_Sigma.py}. These modules have two roles, on one hand  to run the basic functions to compute the initial Fisher matrices in each case; and on the other, to compute the change of variable from the initial Fisher matrices to the desired ones. An example of change of variable is explained in \ref{change}, but this case could be modified to take into account any other set of  variables.

\item \texttt{.\slash FARO\slash Modules\slash Change\_control\slash Basic\_prog\slash}: in this folder we have the modules \texttt{Fisher\_matrices.py}, \texttt{Spec\_cov.py}, \texttt{DP\_param.py}, \texttt{Aux\_fun.py}, \texttt{Win\_dens.py} and \texttt{Functions.py}. Finally, this folder contains the basic \texttt{FARO} functions. The main output of these modules are the initial Fisher matrices for the $A_a(z)$, $R(z)$, $L(z)$ and $E(z)$  parameters in each redshift bin and $a_n$ parameters for the discretization of $\hat{P}(k)$ in each $k$-bin.

\end{itemize}

The code is built to define each quantity in the most general way. To do that, each element is defined as a multidimensional matrix and the operations are made with the \texttt{numpy}  function \texttt{np.einsum} which is very useful to calculate products and sums over indices of multidimensional tensors. This is the reason why  Fisher matrices are defined with finite sums in  (\ref{2.1}), (\ref{2.9}) and (\ref{2.18}), then the observable power spectra are evaluated and summed over a discrete bins for $k$, $\hat{\mu}$, $\ell$ or $z$. For example, the derivative $\left.\frac{\partial P_{a b}^{\delta \delta}(z_i,\hat{\mu}_m,k_c)}{\partial \theta_p}\right|_{r}$ from  the Fisher matrix (\ref{2.1}) is defined in the code as a multidimensional matrix with the structure:

\begin{equation}\label{4.1}
\left.\frac{\partial P_{a b}^{\delta \delta}(z_i,\hat{\mu}_m,k_c)}{\partial \theta_p}\right|_{r} \equiv \texttt{DPT[p][i][m][c][a][b]}.
\end{equation}

This is very useful because if we want to modify the code in a particular point, we just have to calculate numerically or analytically the concrete multidimensional matrix and substitute it in the code. The structure of each matrix is specified in the code. In addition, this approach is useful to compute the inverse covariance matrices of the power spectra, in particular the covariance matrix for the cross correlation power spectrum (\ref{2.19}).

\section{Forecasts for future surveys}\label{secexample}

Finally, we will use $\texttt{FARO}$  with the current specifications of several future galaxy and lensing surveys and compare their sensitivity in different redshift and scale ranges. Thus, in particular, we will focus on  Euclid \cite{Laureijs:2011gra}, DESI \cite{Aghamousa:2016zmz}, J-PAS \cite{Benitez:2014ibt} and LSST \cite{Mandelbaum:2018ouv}. For Euclid we will use the latest specifications from \cite{Blanchard:2019oqi} and analyze separately the photometric and spectroscopic surveys. All the surveys specifications can be found in Appendix \ref{surveyfeatu}. Note that in some cases there are redshift bins in which we only have one or two galaxy tracers. In this situation one should make the multitracer galaxy analysis separately. For example, in DESI we have some redshift bins with only BGS, so we should do on one hand the BGS forecast and on the other hand the LRG+ELG+QSO forecast. Zero values for galaxy densities should never be considered, unless they are in the last redshift bins. For the sake of performing a correct comparison of different galaxy surveys, we have to use redshift bins of the same size. If we used different sizes we would not be comparing the same parameters. In most cases, a larger redshift-bin size implies more information for the parameters in each redshift bin, and smaller errors. However, for the multitracer power spectra, there are correlations between different redshift bins due to (\ref{2.8}) that can increase errors if they are important. For this reason, the redshift bin size should satisfy $\Delta z_i < H_0 \, \chi (z_i) \, E^2(z_i)$ to avoid this effect.
\\
\\
Regarding the $k$ bining, the error in $\hat{P}(k)$ increases with the number of $k$ bins as expected. Correlations between redshift and scale-dependent  parameters are negligible for a reasonable number of $k$ bins. However, if we increase the number of $k$ bins,  these correlations can be relevant and errors for redshift-dependent parameters would increase. As we have checked, the number of $k$-bins should be between $4$ and $20$ according to the volume of the survey. In the examples of this paper, a number of $10$ bins in $k$ is appropriate for all the surveys except for DESI BGS that requires only $5$ bins in $k$ due to the reduced $z$ range.

\subsection{Multitracer 3D galaxy  power spectrum information}

\begin{figure}[H]
\begin{center}
  	\includegraphics[width=0.52\textwidth]{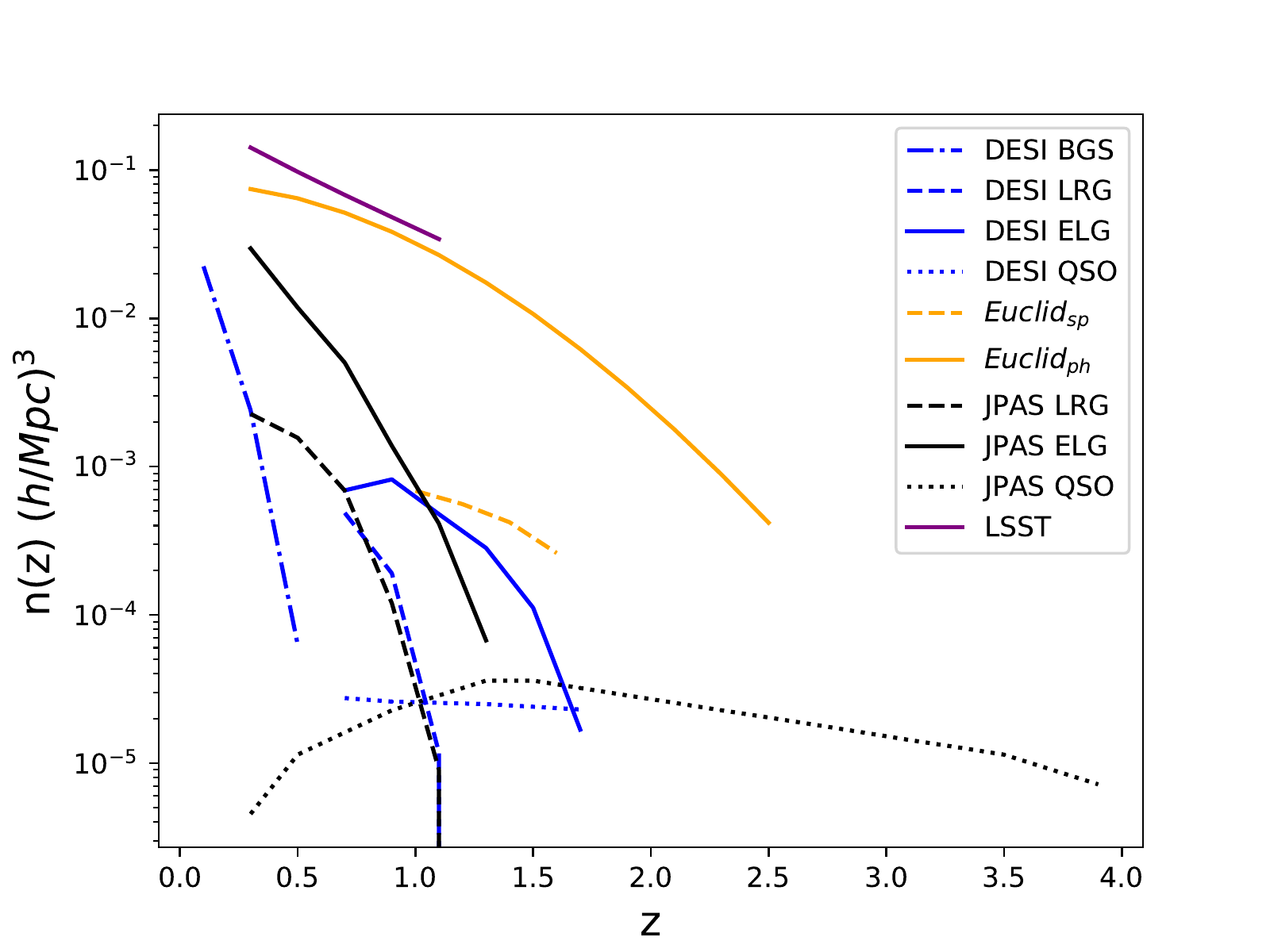}
		\caption{Galaxy densities for each galaxy survey and each tracer in units of $(h/\text{Mpc})^3$ \cite{Resco:2019xve, Aghamousa:2016zmz, Blanchard:2019oqi, Mandelbaum:2018ouv}.}
  \label{Figure_dens}
\end{center}
\end{figure}

Here we analyze the results for the multitracer 3D galaxy power spectrum. In Fig. \ref{Figure_dens} we plot the galaxy densities for each survey and each tracer. 

First we focus on the initial model-independent parameters. In Fig. \ref{Figure_err_1} we plot errors for $E(z)$, $\hat{P}(k)$ and for $R(z)$. As we can see, in order to have good accuracy for $E(z)$ and $R(z)$, it is necessary  to have large number densities of galaxies with precise redshifts, $\delta z^C \lsim 0.01$. So the most appropriate surveys to measure these parameters are spectroscopic and spectro-photometric surveys. On the other hand, for $\hat{P}(k)$, it can be seen that the main characteristic that determines the sensitivity of the survey is the effective volume. In this case, all the surveys analyzed have comparable precision in most of the $k$ range, with more remarkable differences around $k=2\times 10^{-1} \;h/$Mpc. Finally, considering the change of variable with the Planck prior for $\sigma_8$ mentioned before, we plot errors for the growth function $f(z)$ using clustering information in Fig. \ref{Figure_mod_1} left panel.

\begin{figure}[H]
\begin{center}
  	\includegraphics[width=0.52\textwidth]{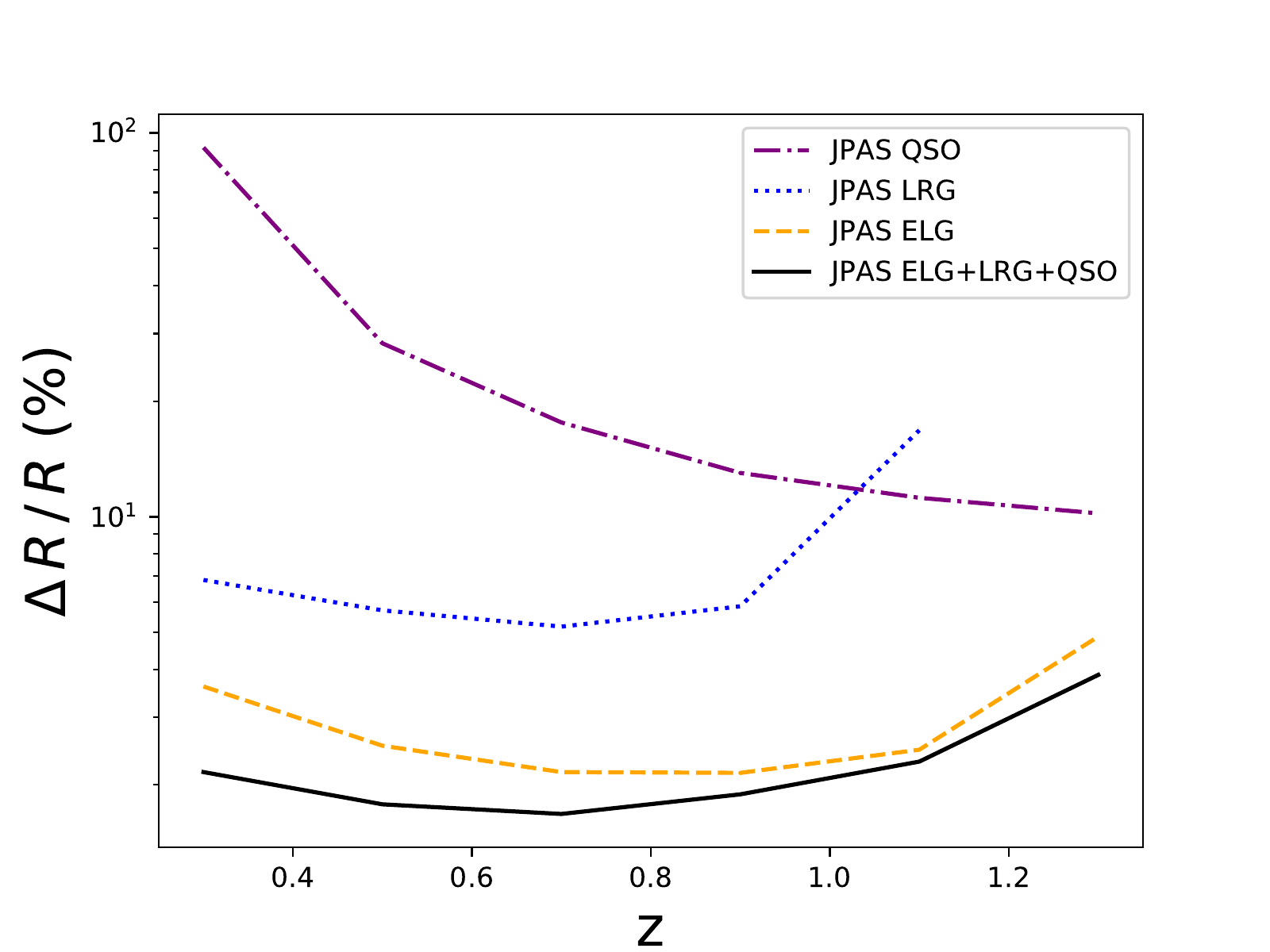}
		\caption{Percentage relative errors for $R(z)$ considering different galaxy tracers of J-PAS survey and the combination of all these tracers.}
  \label{multi}
\end{center}
\end{figure}

Finally, in order to assess the impact of multitracer measurements, in Fig. \ref{multi} we compare the precision in $R(z)$ for the  J-PAS survey using their different galaxy tracers alone and the total combination. As it can be seen, the total combination improves the sensitivity obtained with the best galaxy tracer only that, in this case, are the ELGs. A recent analysis of the improvement from multiple tracer using Fisher formalism can be found in \cite{Boschetti:2020fxr}

%

\subsection{Lensing convergence power spectrum information}

Now we show the results for the lensing convergence power spectrum, focusing on the model-independent parameters. In Fig. \ref{Figure_err_2} we plot errors for $E(z)$, $\hat{P}(k)$ and for $L(z)$. As we can see in the survey specifications, DESI and the spectroscopic Euclid do not collect weak lensing data. In this situation the constraints are essentially dominated by the angular density of galaxies $n_\theta$.

Notice that despite the low redshift precision of the pure photometric
surveys, still good measurements of $E(z)$ are possible using lensing information alone.

\subsection{Multitracer, convergence and cross-correlation power spectra information}

Finally, we combine information from clustering and lensing. First we will focus on the initial model-independent parameters. In Fig. \ref{Figure_err_3} we plot errors for $E(z)$, $\hat{P}(k)$, $R(z)$ and for $L(z)$. The combination of clustering and lensing improves the constraints of both clustering and lensing parameters thanks to the improved constraints on the dimensionless Hubble parameter $E(z)$. The improvement on the Hubble parameter constraints depends on the relative differences of the clustering and lensing sensitivities. In general, if those sensitivities alone are comparable, the combined one improves significantly as is the case of LSST and Euclid$_{ph}$. 

\begin{figure}[H]
\begin{center}
  	\includegraphics[width=0.52\textwidth]{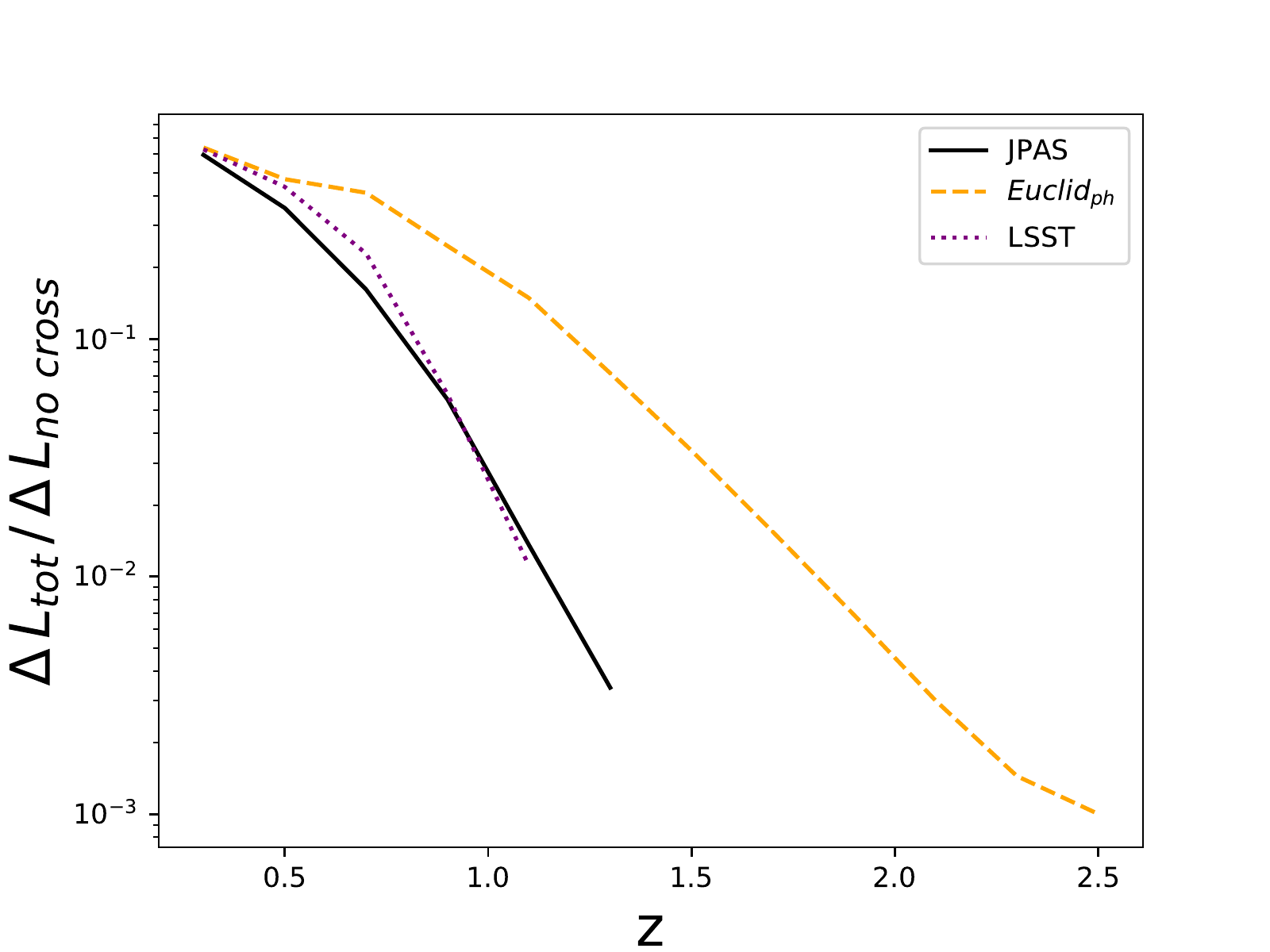}
		\caption{Ratio between errors of $L(z)$ considering clustering, lensing and the cross correlation information, and the errors of $L(z)$ considering clustering and lensing informations without cross correlation. The cross correlation information improves $L(z)$ constraints from one to three orders of magnitude depending on redshift.}
  \label{compare_L}
\end{center}
\end{figure}

Also, as expected if one of the sensitivities is much larger than the other, such sensitivity dominates the combined one. For $\hat{P}(k)$, it can be seen that the combination of clustering and lensing improves the clustering constraints specially at small scales $k\gsim 10^{-1}\;h/$Mpc.
\\
\\
Finally, considering the change of variable with the Planck prior for $\sigma_8$ mentioned before, we plot in Fig. \ref{Figure_mod_1} right panel the errors for $f(z)$ and in Fig. \ref{Figure_mod_2} the errors for $\Sigma (z)$. Notice that one of the main advantages of combining clustering and lensing, and using a prior on $\sigma_8$, is that it is possible to measure $\Sigma (z)$ in each redshift bin.

\subsection*{The impact of cross-correlation}

\begin{figure}[H]
\begin{center}
  	\includegraphics[width=0.52\textwidth]{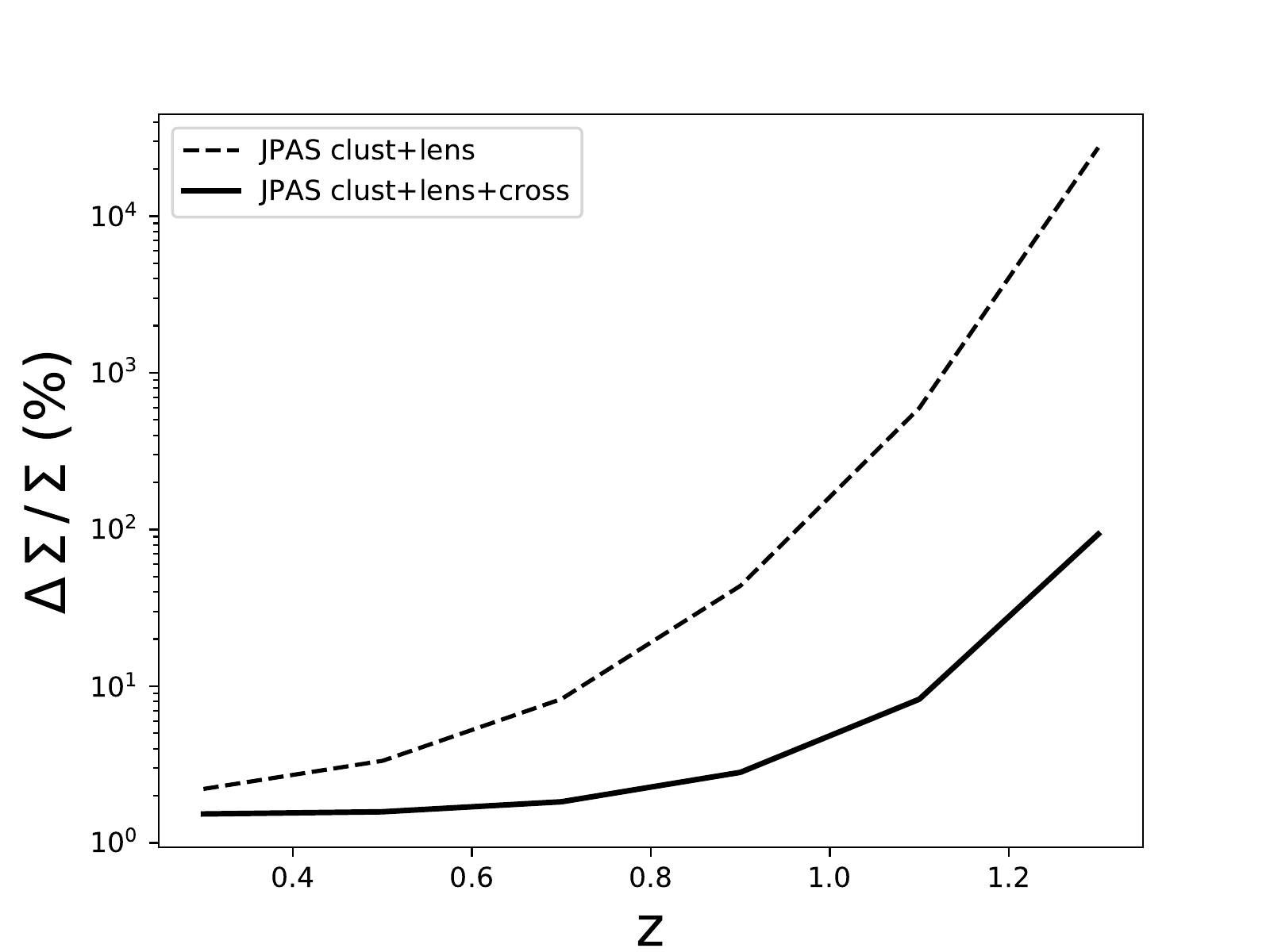}
		\caption{Relative errors for $\Sigma(z)$ using J-PAS survey with clustering and lensing information; and with clustering, lensing and the cross correlation information.}
  \label{Figure_Sigma_no_cross}
\end{center}
\end{figure}

In order to estimate the impact of the cross-correlation
between clustering and lensing, in Fig. \ref{compare_L}
we plot the ratio between the precision of the lensing parameter $L(z)$ obtained with and without cross-correlation. As it can be seen the cross-correlation can improve the precision, up to three orders of magnitude
depending on redshift. Similarly, the constraints on $\Sigma (z)$ obtained with the $\sigma_8$ prior depends strongly on both clustering and lensing information. In particular, as it can be seen in Fig. \ref{Figure_Sigma_no_cross}, the cross-correlation information between clustering and lensing improves significantly the precision. The relevant
role of cross-correlation in the determination of certain
parameters has been recently discussed for Euclid$_\text{ph}$  in \cite{Tutusaus:2020xmc}.

This analysis has been done considering the different data of all future surveys separately. 
Thus, each galaxy survey has its best targets and redshift ranges. A combined analysis of the  data
collected by different surveys would give  more precise cosmological measurements. In particular, the combination of spectroscopic and photometric surveys improves significantly the constraints, as can be seen with a spectro-photometric survey like J-PAS.

\section{Conclusions}\label{con}

We have presented the new public Python $\texttt{FARO}$ code which has been designed to perform model-independent  Fisher forecast analysis for multitracer galaxy and lensing surveys. The code has been written to be as fast and versatile as possible. It is based on multidimensional matrices for each mathematical element so that they can be modified and replaced in a simple way. These multidimensional matrices are manipulated and computed using the useful Python function $\texttt{np.einsum}$. The main observables used by $\texttt{FARO}$ are the multitracer 3D power spectrum, the lensing convergence power spectrum and the power spectrum for the multitracer cross correlation between galaxy distribution and shapes. With these observables, we are able to forecast the sensitivity of the main future galaxy surveys like Euclid, DESI, JPAS or LSST. The code follows a model-independent approach in which we consider as free parameters $A_a(z)$, $R(z)$, $L(z)$ and $E(z)$ in each redshift bin, in addition to a model-independent parametrization of $\hat{P}(k)$ in logarithmically spaced $k$ bins. A variable change module is provided which allows to project the Fisher matrices into any desired set of new parameters. An example of change of variable to parameters $b_a(z)$, $f(z)$, $\Sigma(z)$ and $E(z)$ in each redshift bin is explicitly included in the code. As can be seen in our examples, the combination of clustering, lensing and the cross-correlation information improves significantly the constraints. This imply that a survey with a modest number of galaxies that combines these
three observables can reach a precision comparable to the precision of a larger survey with larger  number of galaxies but focused in one type of observable. Some of the main utilities of $\texttt{FARO}$ are: analyzing the sensitivity of different surveys in the different redshift and scale ranges; analyzing how the use of different tracers, combination of different observables or the various observation strategies can affect the sensitivity of a survey; or analyzing which is the set of parameters that can be measured with the best precision depending on the specifications of a given survey.

As compared to other Fisher matrix codes for galaxy surveys, $\texttt{FARO}$ follows a simple approach in which it is possible to compute derivatives analytically. This approach has only two important approximations: flat FRW background metric and scale-independent growth of perturbations. Considering these features, the code is faster than standard codes in which derivatives are computed numerically. This makes the code a useful tool to analyze multiple survey configurations with different observables and tracers.

{\bf \texttt{FARO} availability}: the code is publicly available from \texttt{https://www.ucm.es/iparcos/faro}.

\vspace{0.2cm}
{\bf Acknowledgements:}
We thank the J-PAS Theoretical Cosmology and Fundamental Physics working group for useful comments and discussions.  M.A.R acknowledges support from UCM predoctoral grant.  This work has been supported by the MINECO (Spain) project FIS2016-78859-P(AEI/FEDER, UE).


\clearpage

\begin{widetext}

\section{Appendix A: $\texttt{FARO}$ fluxchart.}

In this appendix we plot the fluxchart of the $\texttt{FARO}$ structure. Arrows from a module indicate which modules are used in such module. The module color refers to the folder in which the module is located. As it can be seen, when $\texttt{FARO.py}$ is run all modules are called.

\begin{figure} [h!]
  	\includegraphics[width=0.95\textwidth]{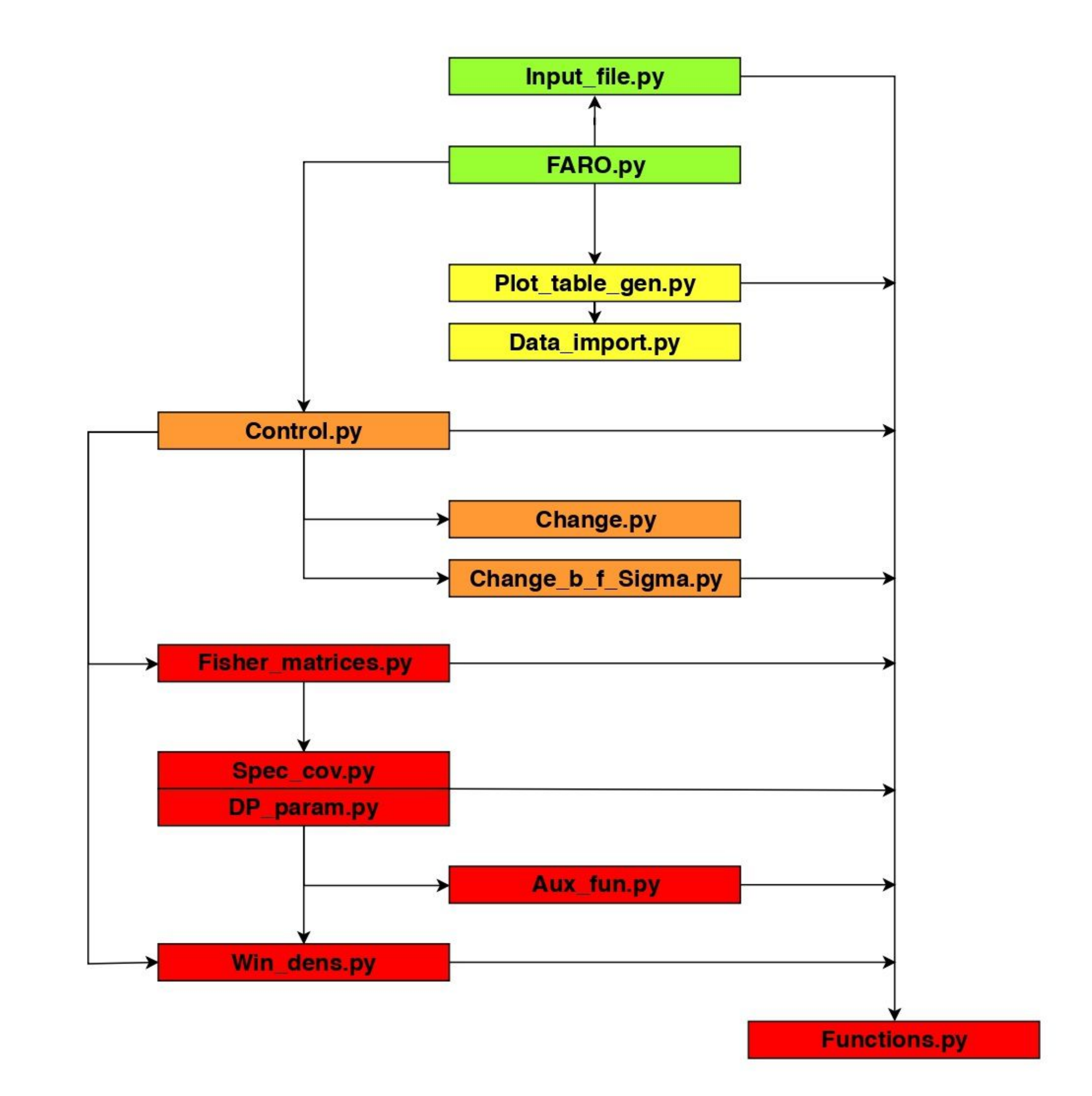}
		\caption{Fluxchart of the structure of $\texttt{FARO}$. In green the modules that are contained in $\texttt{.\slash FARO\slash}$. In yellow the modules that are contained in $\texttt{.\slash FARO\slash Modules\slash}$. In orange the modules that are contained in $\texttt{.\slash FARO\slash Modules\slash Change\_control\slash}$. Finally, in red the modules that are contained in $\texttt{.\slash FARO\slash Modules\slash Change\_control\slash Basic\_prog\slash}$.}
  \label{Figure_1}
\end{figure}

\newpage

\section{Appendix B: survey specifications.}\label{surveyfeatu}

Now we summarize all the inputs for the examples of section \ref{secexample}. The fiducial cosmology we use for all the forecast has the following values: $\Omega_m=0.31$, $\omega_{b}=0.0226$, $n_{s}=0.96$, $h=0.68$, $H^{-1}_{0}=2997.9 \, \textrm{Mpc/h}$ and $\sigma_{8}=0.82$. For the growth function we use the $\mathrm{\Lambda CDM}$ expression $f(z) = \Omega_m (a)^{\gamma}$ with $\gamma = 0.545$ \cite{Linder:2007hg}. In addition, for all the surveys we use the same values for the non-linear cut-off $\Sigma_0 = 11 \, \mathrm{Mpc/}h$ and the intrinsic ellipticity $\gamma_{int} = 0.22$. Redshift bin centers of each survey are shown in Table \ref{taJPASDESI}. Scale bins are logspaced in $10$ bins from $k = 0.007 \, h/\mathrm{Mpc}$ to $k = 1 \, h/\mathrm{Mpc}$. The sky area, redshift errors for clustering and lensing, and the areal galaxy densities can be found in Table \ref{table_spe}. For the bias, we consider four different types of galaxies: Luminous Red Galaxies (LRGs), Emission Line Galaxies (ELGs), Bright Galaxies (BGS) and quasars (QSO) \cite{Mostek:2012nc, Ross:2009sn}. Each type has different fiducial bias given by
\begin{equation}\label{28}
b(z)=\frac{b_0}{D(z)},
\end{equation}
being $b_{0}=0.84$ for ELGs, $b_{0}=1.7$ for LRGs and $b_{0}=1.34$ for BGS. For photometric Euclid survey we use a fiducial bias for ELGs of the form $b(z)=\sqrt{1+z}$ \cite{Laureijs:2011gra}, for spectroscopic Euclid, the fiducial bias values are shown in Table \ref{taJPASDESI}, while the bias for quasars is $b(z)=0.53+0.289 \, (1+z)^{2}$. Finally the bias of LSST galaxies follows equation (\ref{28}) with $b_{0}=0.95$ \cite{Mandelbaum:2018ouv}. For lensing, galaxy distribution is given by (\ref{1.14}) with the following values of the mean redshift: $z_{mean}=0.5$ for J-PAS and $z_{mean}=0.9$ for Euclid, for LSST the galaxy density distribution is, following \cite{Mandelbaum:2018ouv}, $n(z) \propto z^{2} \, \mathrm{exp}\left[-(z/0.11)^{0.68} \right]$ which is used instead of (\ref{1.14}).
\begin{table}[H]
\begin{center}
\resizebox{6.3cm}{!} {\begin{tabular}{c|c|c|c|c|}
\cline{2-5}
\multicolumn{1}{l|}{}  & \multicolumn{1}{c|}{sky area ($\mathrm{deg}^2$)} & \multicolumn{1}{c|}{$\delta z^C$} & \multicolumn{1}{c|}{$\delta z^L$} & \multicolumn{1}{c|}{$n_{\theta}$} \\ \hline
\multicolumn{1}{|c|}{J-PAS} & 8500 & 0.003 & 0.03 & 12 \\ \hline
\multicolumn{1}{|c|}{DESI} & 14000 & 0.0005 & - & - \\ \hline
\multicolumn{1}{|c|}{LSST} & 14300 & 0.03 & 0.05 & 27 \\ \hline
\multicolumn{1}{|c|}{$\mathrm{Euclid_{sp}}$} & 15000 & 0.001 & - & - \\ \hline
\multicolumn{1}{|c|}{$\mathrm{Euclid_{ph}}$} & 15000 & 0.05 & 0.05 & 30 \\ \hline
\end{tabular}}
\end{center}
\caption{Sky areas, redshift errors for clustering $\delta z^C$ and lensing $\delta_z^L$ and the angular galaxy density $n_{\theta}$ 
in galaxies per square arc min. for each survey.}\label{table_spe}
\end{table}
\vspace{-0.3cm}
\begin{table*}[htbp]
\begin{center}
\begin{tabular}{|c|c|c|c|}
\hline
\multicolumn{4}{|c|}{J-PAS} \\ \hline
\hline
$z$ & $LRG$ & $ELG$ & $QSO$ \\
\hline \hline
0.3 & 226.6 & 2958.6 & 0.45 \\ \hline
0.5 & 156.3 & 1181.1 & 1.14 \\ \hline
0.7 & 68.8  & 502.1  & 1.61 \\ \hline
0.9 & 12.0  & 138.0  & 2.27 \\ \hline
1.1 & 0.9   & 41.2   & 2.86 \\ \hline
1.3 & 0     & 6.7    & 3.60 \\ \hline
1.5 & 0     & 0      & 3.60 \\ \hline
1.7 & 0     & 0      & 3.21 \\ \hline
1.9 & 0     & 0      & 2.86 \\ \hline
2.1 & 0     & 0      & 2.55 \\ \hline
2.3 & 0     & 0      & 2.27 \\ \hline
2.5 & 0     & 0      & 2.03 \\ \hline
2.7 & 0     & 0      & 1.81 \\ \hline
2.9 & 0     & 0      & 1.61 \\ \hline
3.1 & 0     & 0      & 1.43 \\ \hline
3.3 & 0     & 0      & 1.28 \\ \hline
3.5 & 0     & 0      & 1.14 \\ \hline
3.7 & 0     & 0      & 0.91 \\ \hline
3.9 & 0     & 0      & 0.72 \\ \hline
\end{tabular}\hspace{0.3in}\begin{tabular}{|c|c|}
\hline
\multicolumn{2}{|c|}{$\mathrm{Euclid_{ph}}$} \\ \hline
\hline
$z$ & $ELG$ \\
\hline \hline
0.3 & 7440  \\ \hline
0.5 & 6440  \\ \hline
0.7 & 5150  \\ \hline
0.9 & 3830  \\ \hline
1.1 & 2670  \\ \hline
1.3 & 1740  \\ \hline
1.5 & 1070  \\ \hline
1.7 & 620   \\ \hline
1.9 & 341   \\ \hline
2.1 & 178   \\ \hline
2.3 & 88.3  \\ \hline
2.5 & 41.8  \\ \hline
\end{tabular}\hspace{0.3in}\begin{tabular}{|c|c|c|c|c|}
\hline
\multicolumn{5}{|c|}{DESI} \\ \hline
\hline
$z$ & $BGS$ & $LRG$ & $ELG$ & $QSO$ \\
\hline \hline
0.1 & 2240 & 0    & 0    & 0      \\ \hline
0.3 & 240  & 0    & 0    & 0      \\ \hline
0.5 & 6.3  & 0    & 0    & 0      \\ \hline
0.7 & 0    & 48.7 & 69.1 & 2.75   \\ \hline
0.9 & 0    & 19.1 & 81.9 & 2.60   \\ \hline
1.1 & 0    & 1.18 & 47.7 & 2.55   \\ \hline
1.3 & 0    & 0    & 28.2 & 2.50   \\ \hline
1.5 & 0    & 0    & 11.2 & 2.40   \\ \hline
1.7 & 0    & 0    & 1.68 & 2.30   \\ \hline
\end{tabular}\hspace{0.3in}\begin{tabular}{|c|c|}
\hline
\multicolumn{2}{|c|}{LSST} \\ \hline
\hline
$z$ & $n$ \\
\hline \hline
0.3 & 14170 \\ \hline
0.5 & 9720  \\ \hline
0.7 & 6790  \\ \hline
0.9 & 4810  \\ \hline
1.1 & 3420  \\ \hline
\end{tabular}\hspace{0.3in}\begin{tabular}{|c|c|c|}
\hline
\multicolumn{3}{|c|}{$\mathrm{Euclid_{sp}}$} \\ \hline
\hline
$z$ & $ELG$ & $b(z)$ \\
\hline \hline
1.0 & 68.6 & 1.46  \\ \hline
1.2 & 55.8 & 1.61  \\ \hline
1.4 & 42.1 & 1.75  \\ \hline
1.6 & 26.1 & 1.90  \\ \hline
\end{tabular}
\caption{From left to right: redshift bins and densities of luminous red galaxies (LRG), emission line galaxies (ELG) and quasars (QSO) for J-PAS. Redshift bins and densities of ELG for photometric Euclid survey. Redshift bins and densities of bright galaxies (BGS), LRG, ELG and QSO for DESI. Redshift bins and galaxy densities for LSST,  and finally, redshift bins, emission line galaxy densities and bias for spectroscopic Euclid survey. Galaxy densities in units of $10^{-5}$ $h^3\mathrm{ \, Mpc^{-3}}$.}
\label{taJPASDESI}
\end{center}
\end{table*}

\newpage

\section{Appendix C: $\texttt{FARO}$ examples.}

In this appendix we show the plots of errors in \ref{secexample} for the model-independent parameters (\ref{1.1}-\ref{1.4}) and (\ref{s1.18}).

\begin{center}
    \subsection{CLUSTERING INFORMATION:}
\end{center}

\begin{figure}[H]
\begin{center}
  	\includegraphics[width=0.495\textwidth]{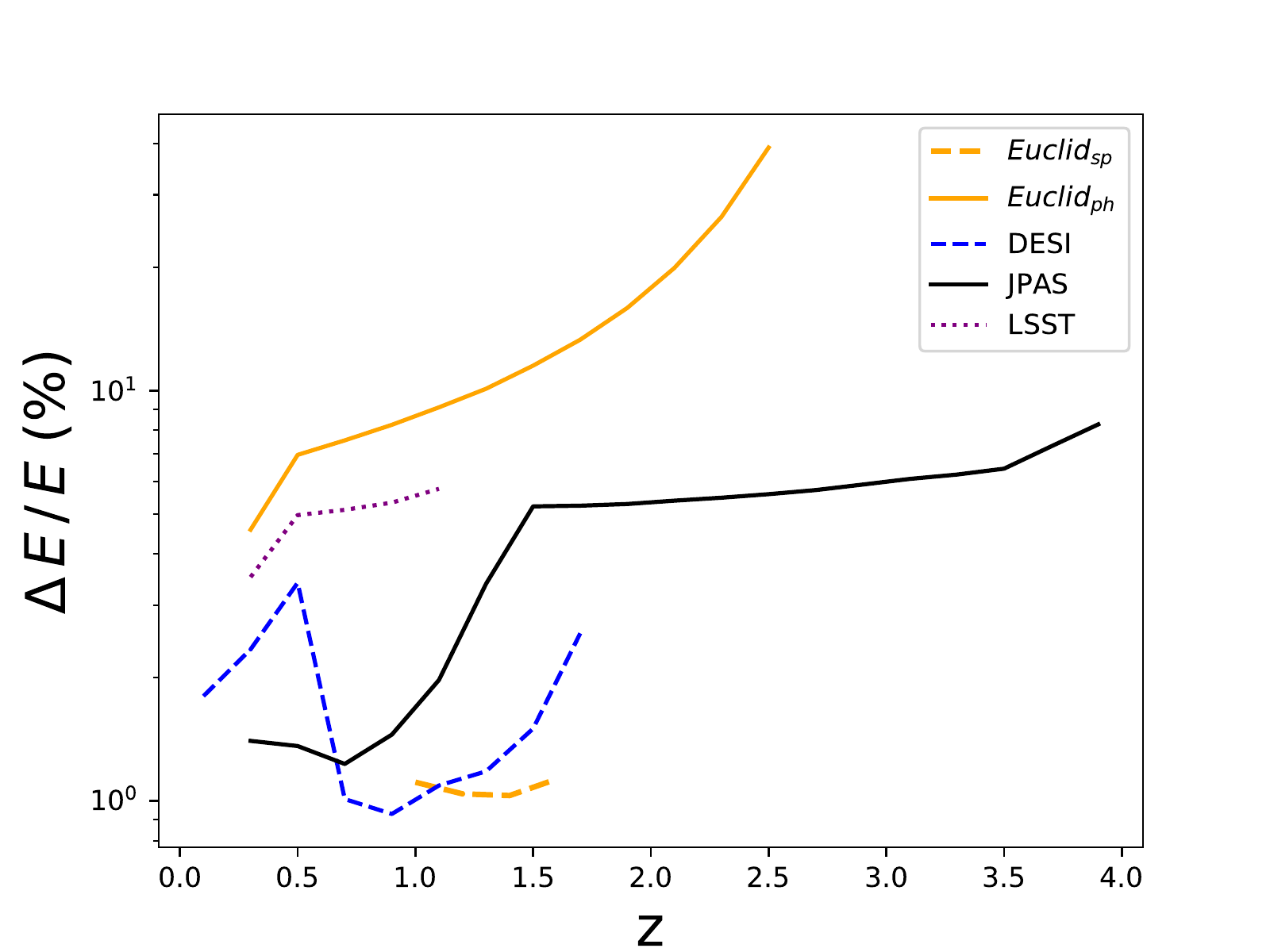}
  	\includegraphics[width=0.495\textwidth]{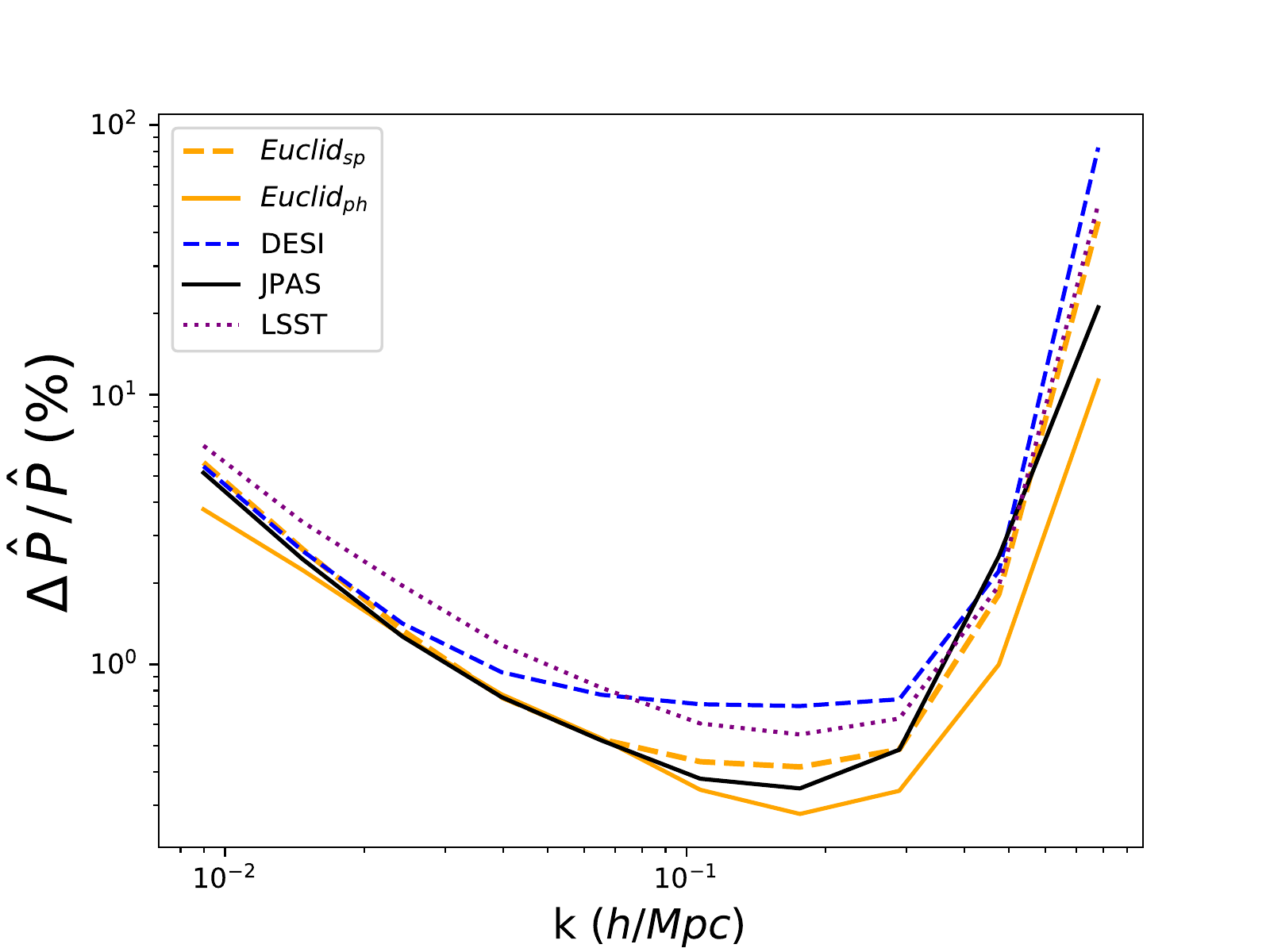}
  	\includegraphics[width=0.495\textwidth]{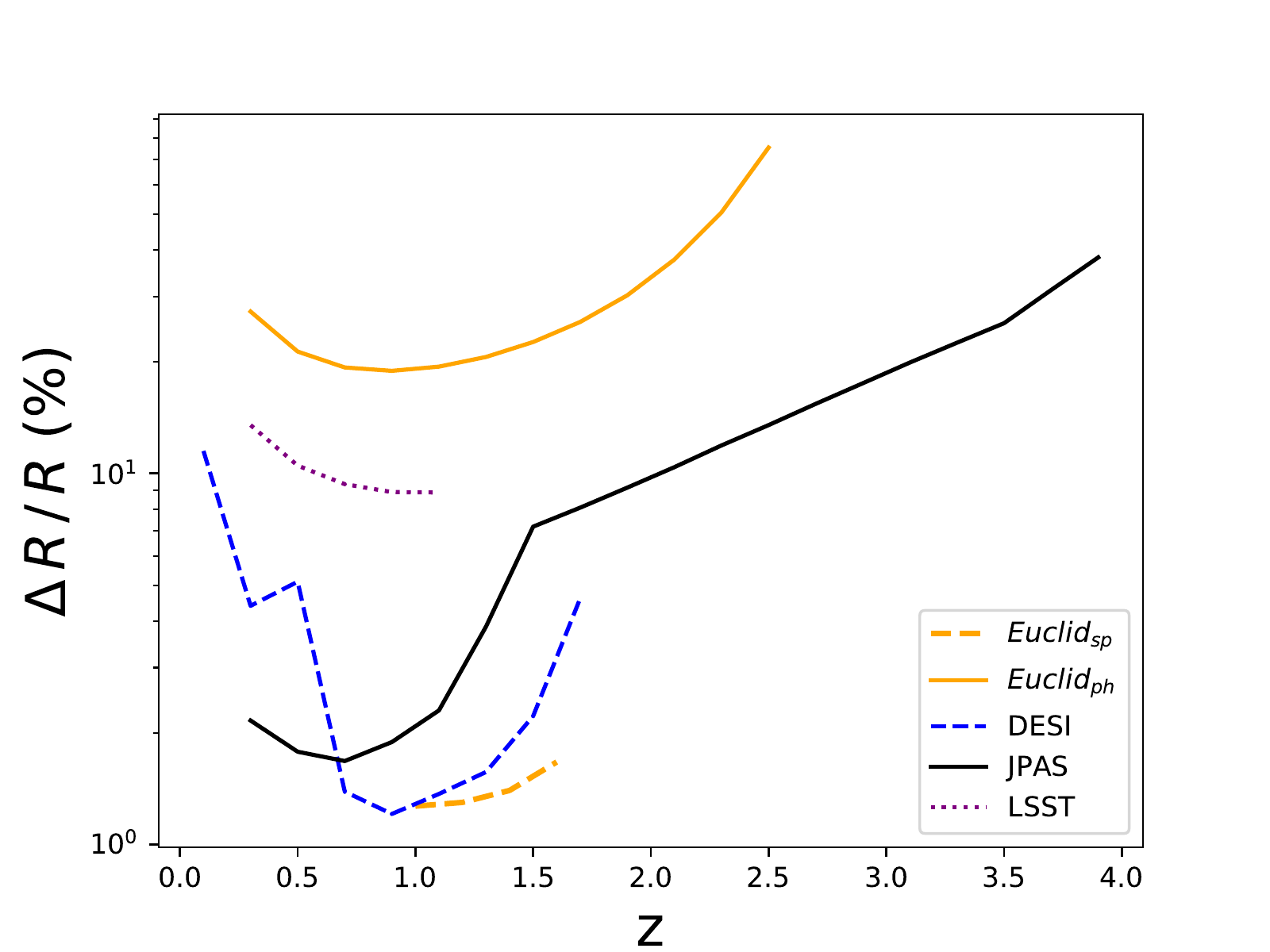}
		\caption{Left up: percentage relative errors for $E(z)$ in each redshift bin for different galaxy surveys using clustering information. Right up: percentage relative errors for $\hat{P}(k)$ in each $k$ bin for different galaxy surveys using clustering information. Down: Percentage relative errors for $R(z)$ in each redshift bin for different galaxy surveys using clustering information.}
  \label{Figure_err_1}
\end{center}
\end{figure}

\newpage

\begin{center}
    \subsection{LENSING INFORMATION:}
\end{center}

\begin{figure}[H]
\begin{center}
  	\includegraphics[width=0.495\textwidth]{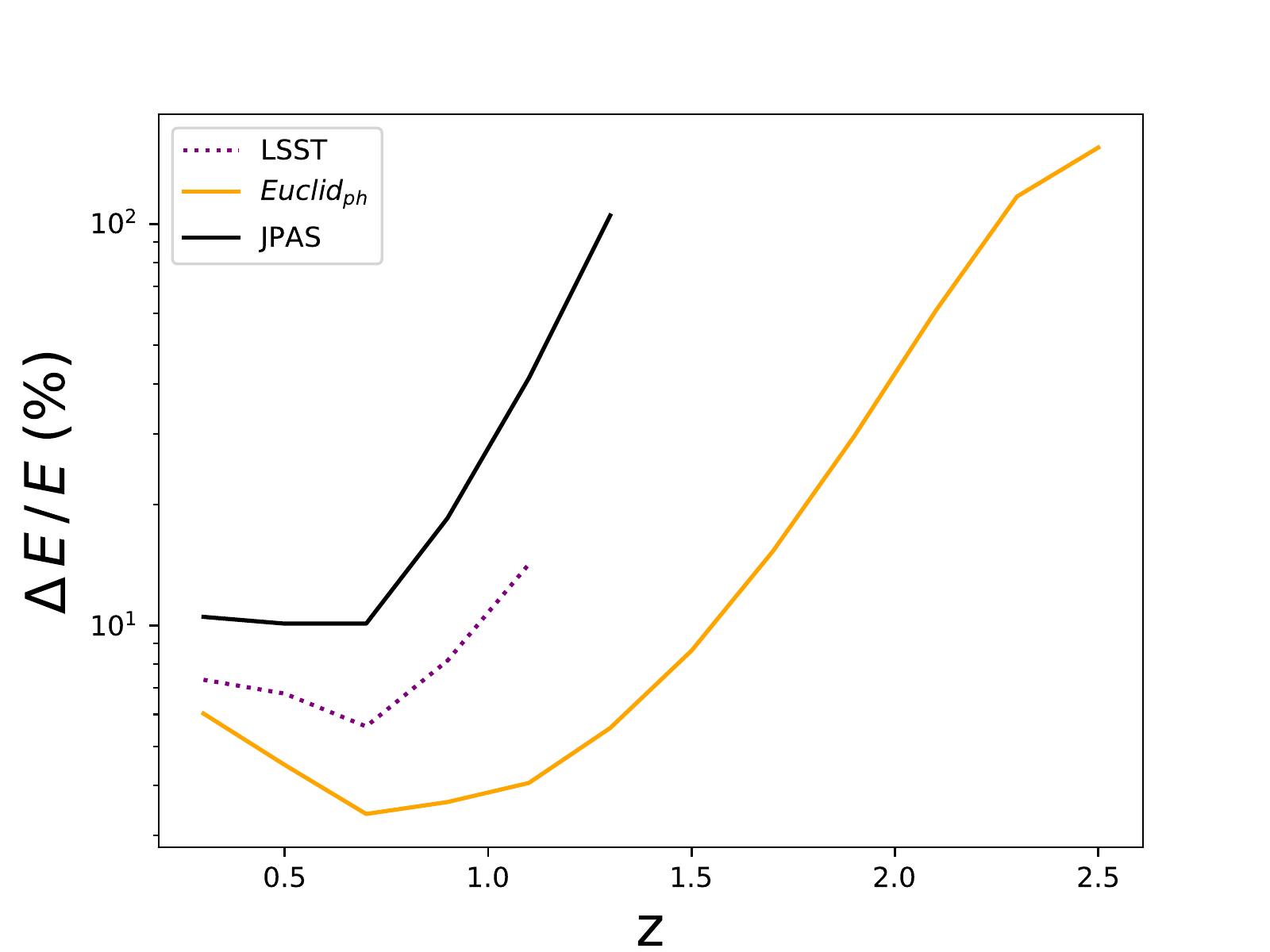}
  	\includegraphics[width=0.495\textwidth]{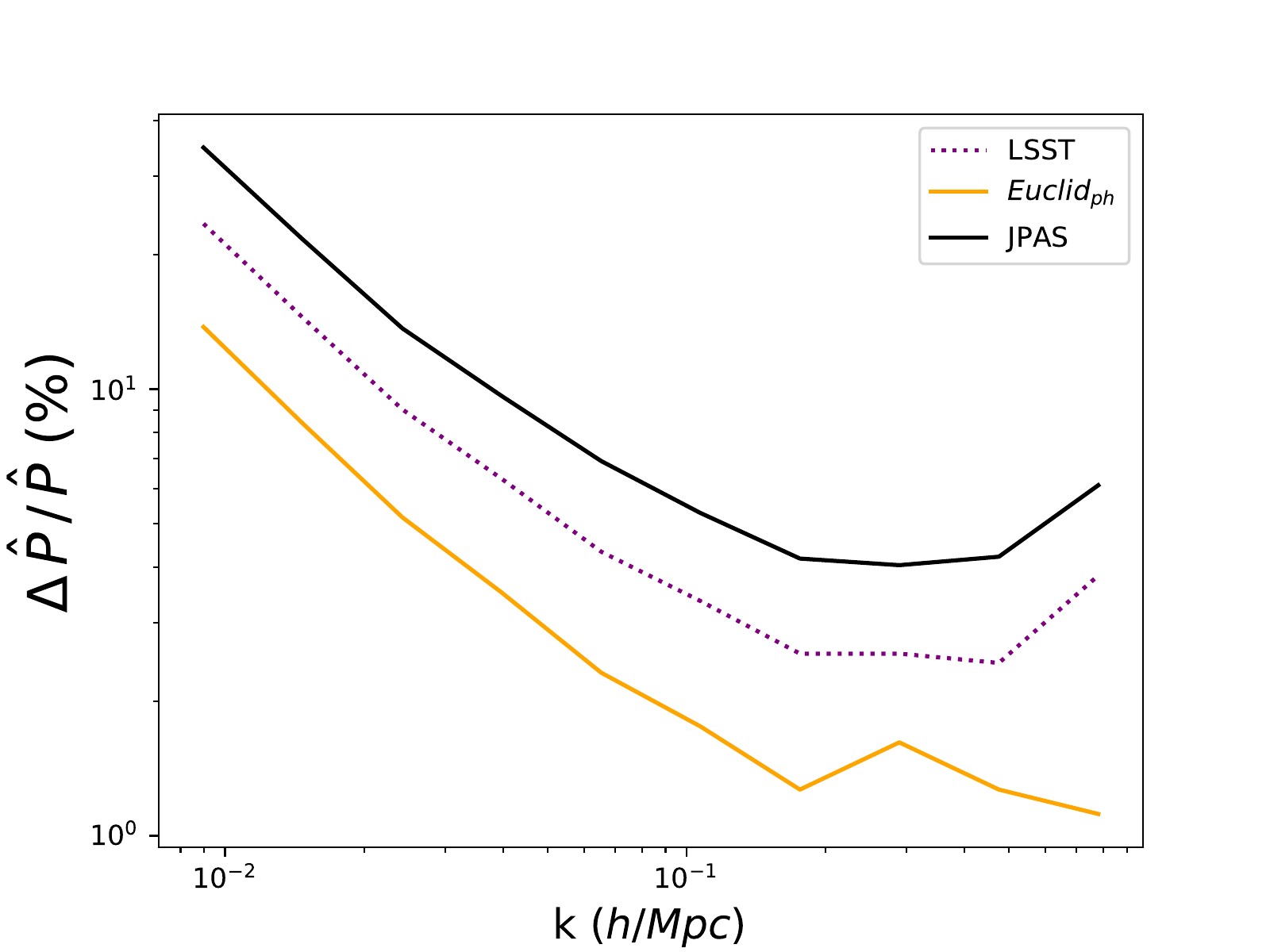}
  	\includegraphics[width=0.495\textwidth]{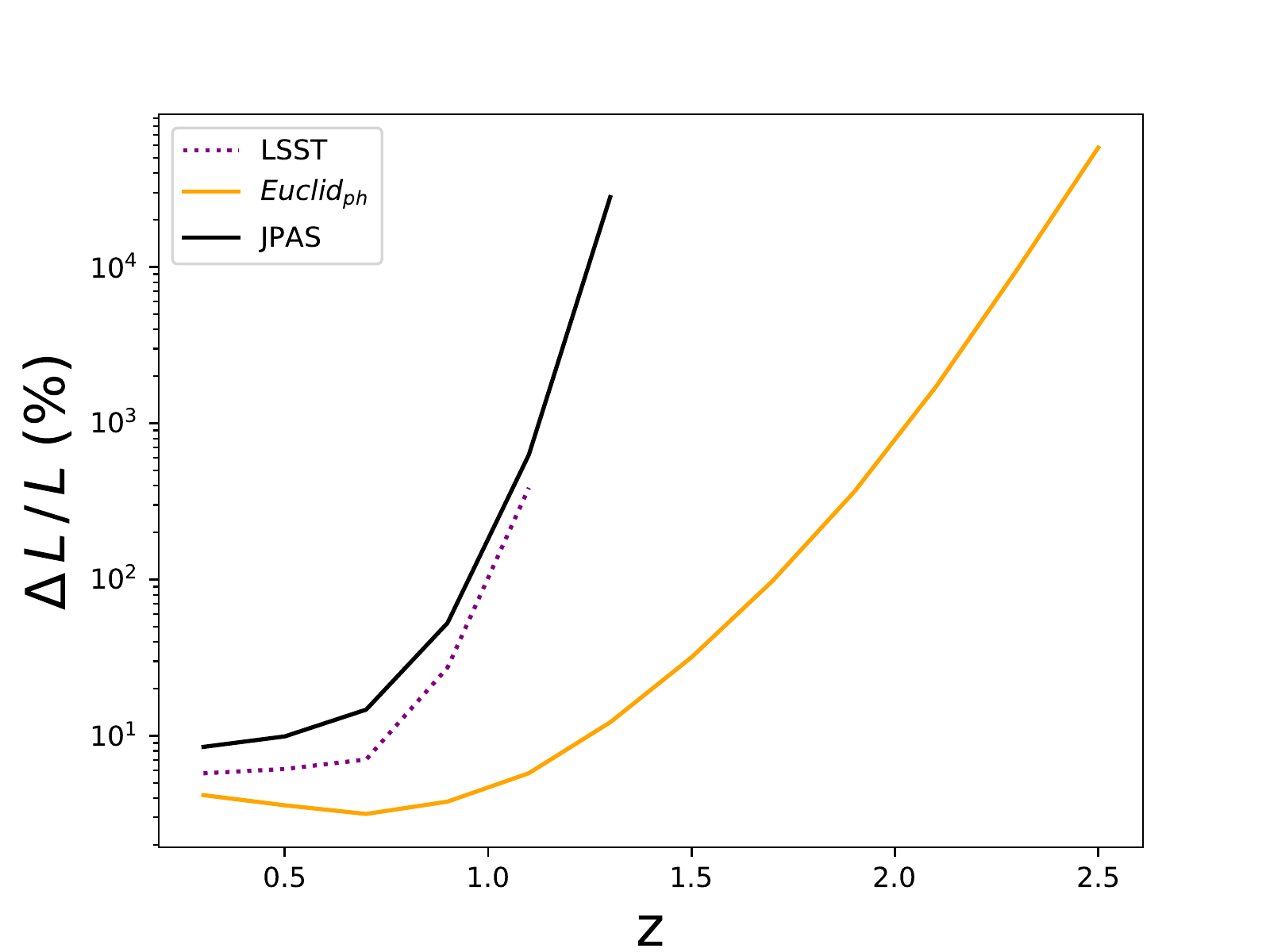}
		\caption{Left up: percentage relative errors for $E(z)$ in each redshift bin for different galaxy surveys using lensing information. Right up: percentage relative errors for $\hat{P}(k)$ in each $k$ bin for different galaxy surveys using lensing information. Down: Percentage relative errors for $L(z)$ in each redshift bin for different galaxy surveys using lensing information.}
  \label{Figure_err_2}
\end{center}
\end{figure}

\newpage

\begin{center}
    \subsection{CLUSTERING + LENSING INFORMATION:}
\end{center}

\begin{figure}[H]
  	\includegraphics[width=0.495\textwidth]{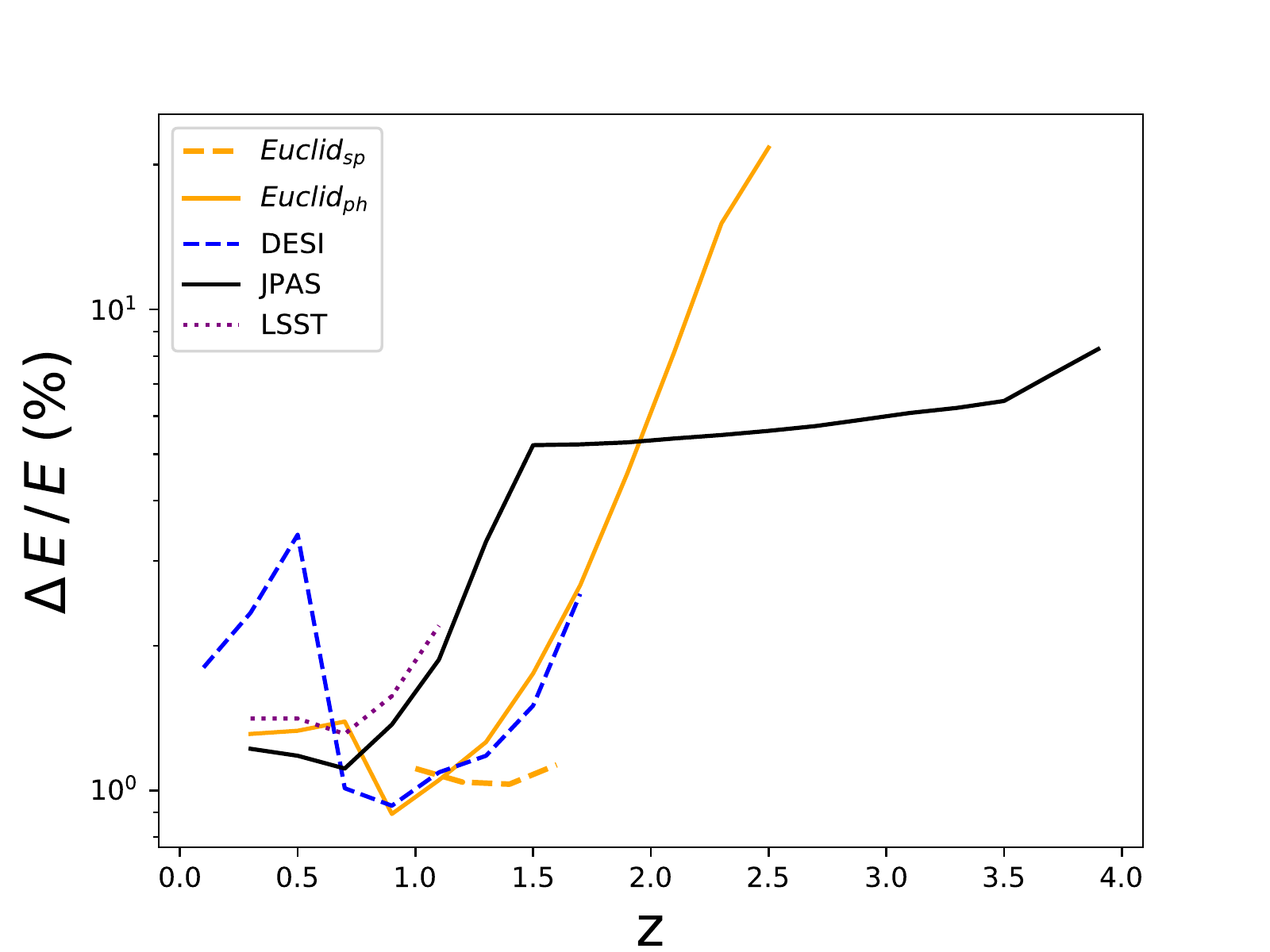}
  	\includegraphics[width=0.495\textwidth]{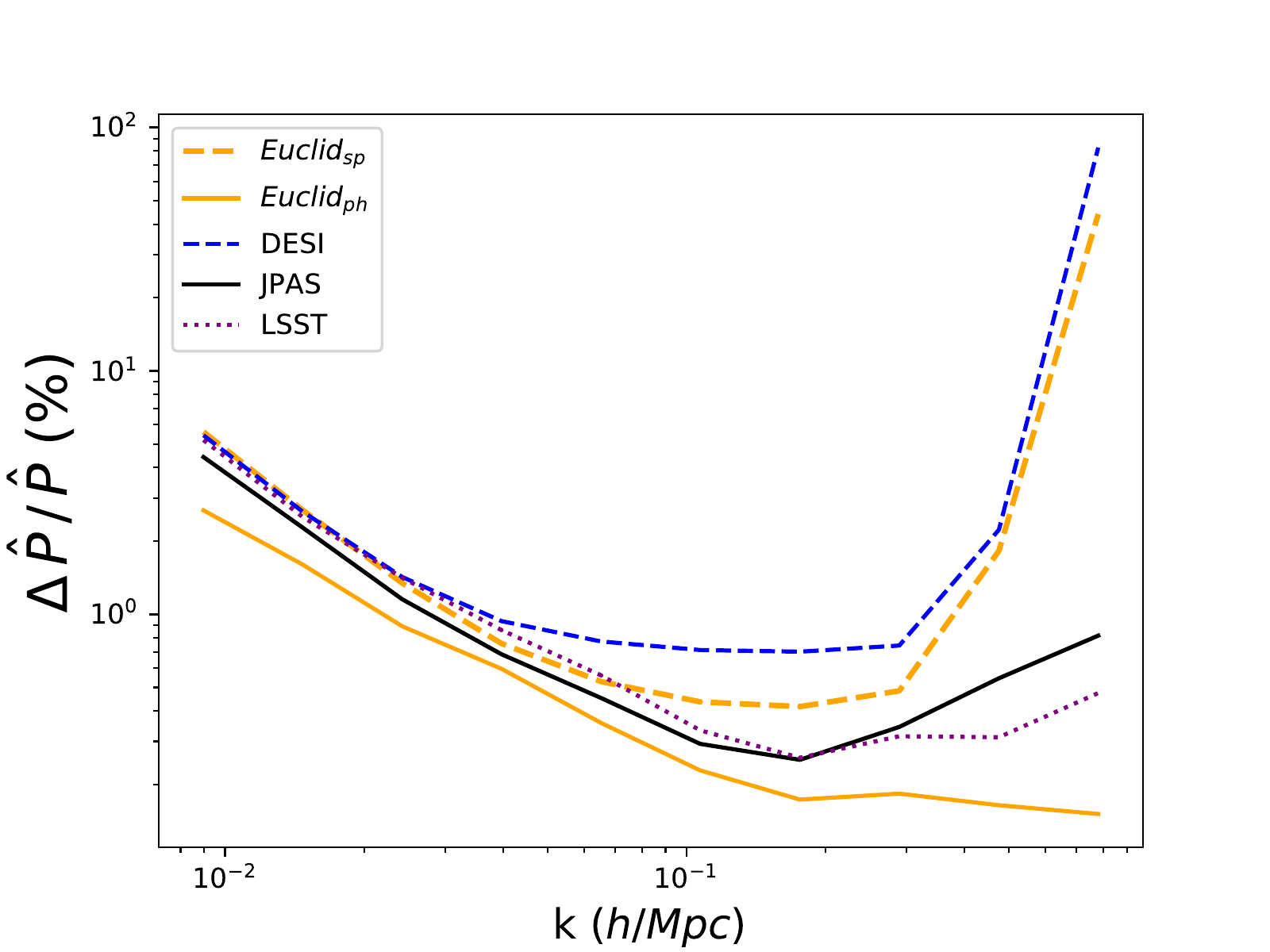}
  	\includegraphics[width=0.495\textwidth]{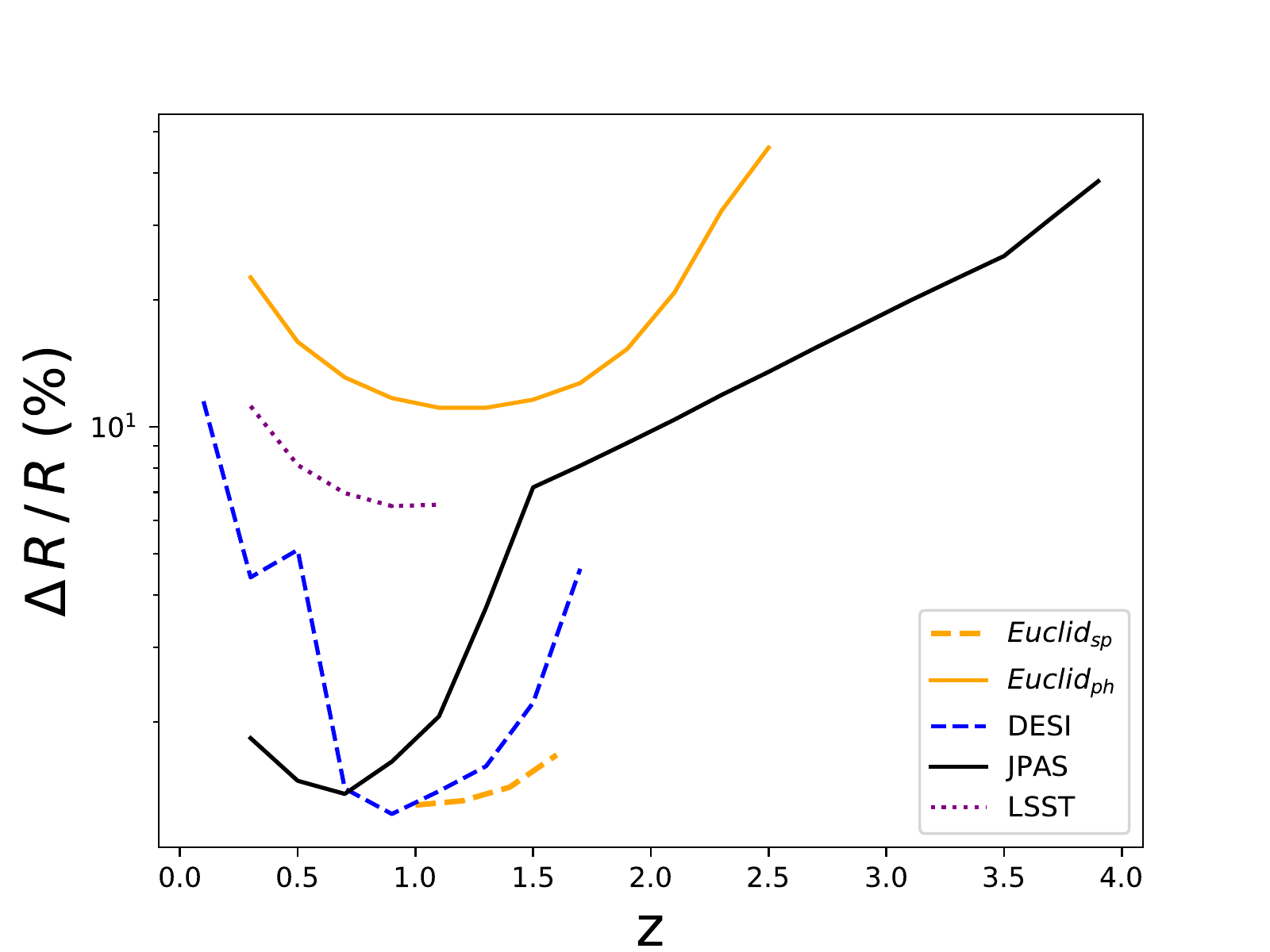}
  	\includegraphics[width=0.495\textwidth]{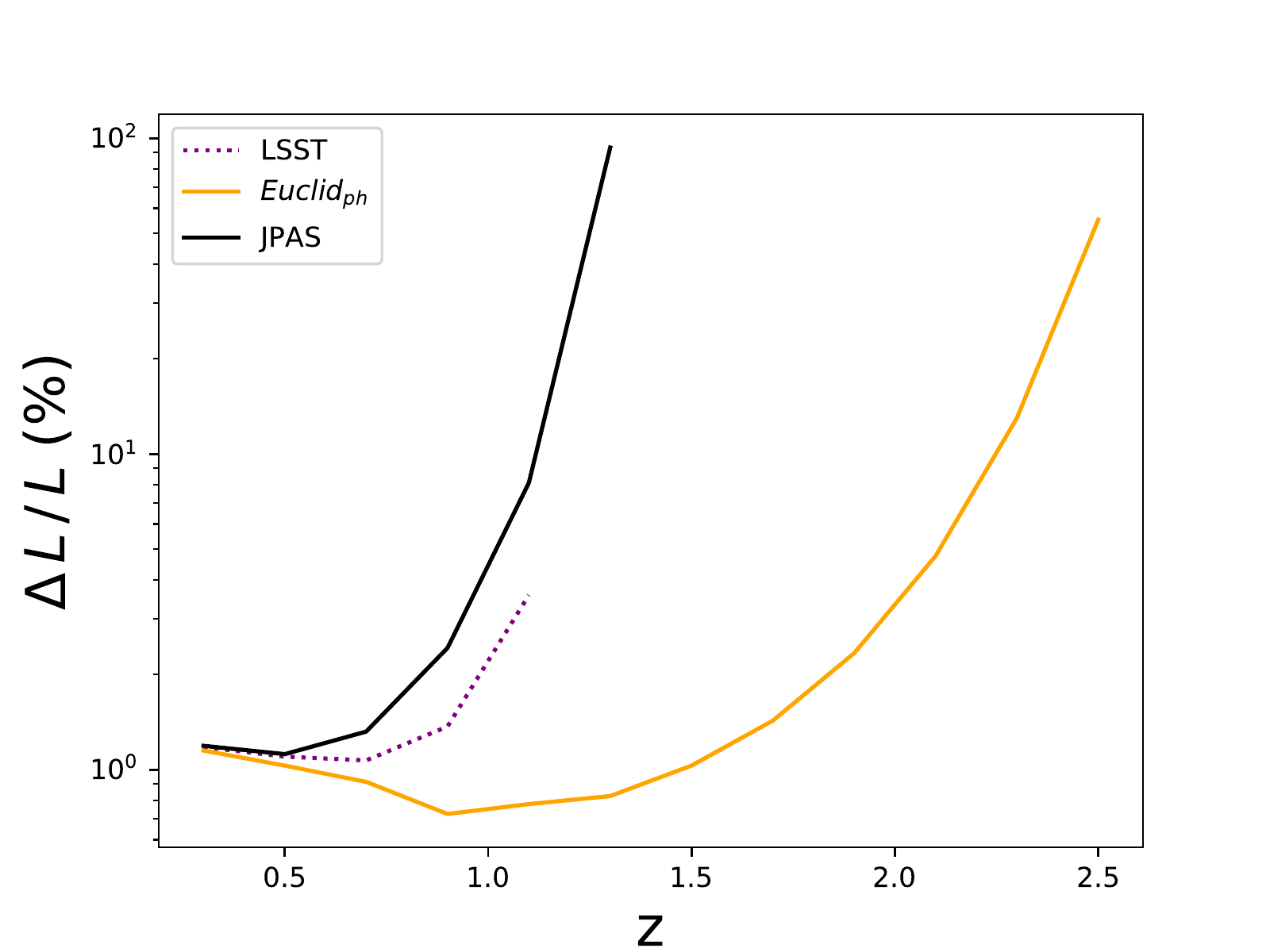}
		\caption{Left up: percentage relative errors for $E(z)$ in each redshift bin for different galaxy surveys using clustering, lensing and the cross correlation information. Right up: percentage relative errors for $\hat{P}(k)$ in each $k$ bin for different galaxy surveys using clustering, lensing and the cross correlation information. Left down: percentage relative errors for $R(z)$ in each redshift bin for different galaxy surveys using clustering, lensing and the cross correlation information. Right down: percentage relative errors for $L(z)$ in each redshift bin for different galaxy surveys using clustering, lensing and the cross correlation information.}
  \label{Figure_err_3}
\end{figure}
%


\newpage

\section{Appendix D: $\texttt{FARO}$ constraints for $f(z)$ and $\Sigma(z)$ using priors.}

In this appendix we show the plots of errors in \ref{secexample} for the modified gravity case \ref{mod}.

\begin{figure} [h!]
  	\includegraphics[width=0.495\textwidth]{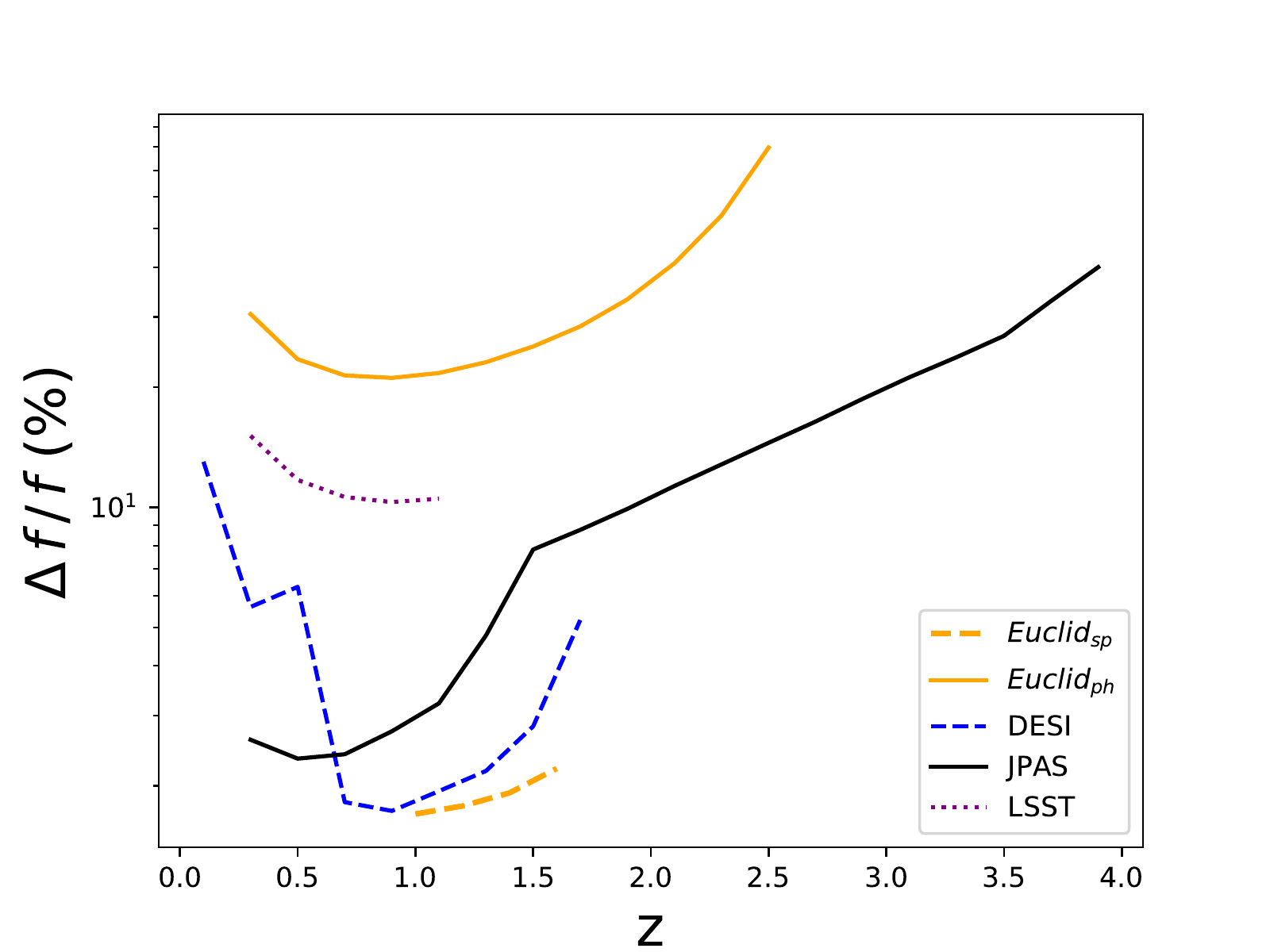}
  	\includegraphics[width=0.495\textwidth]{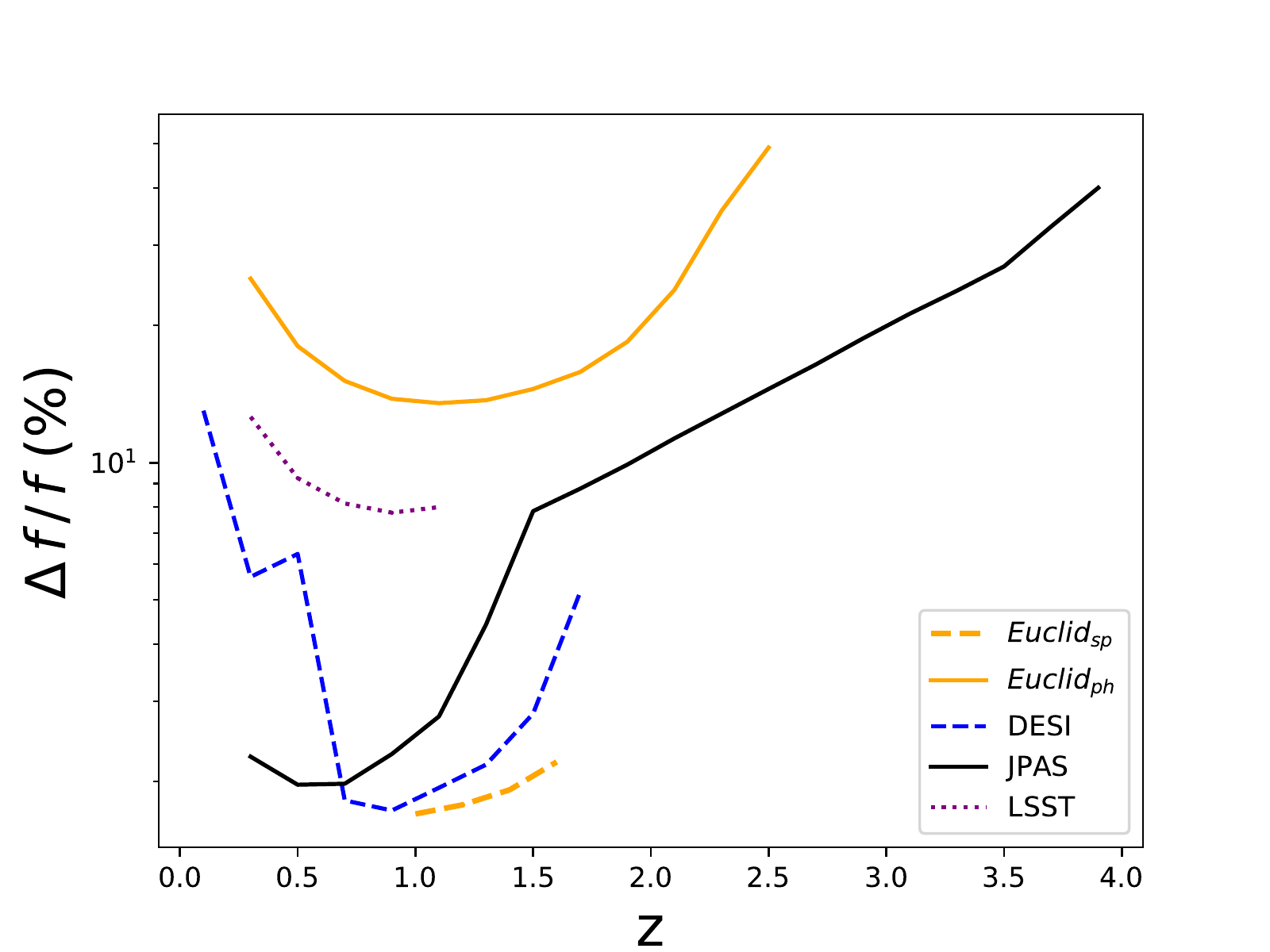}
		\caption{Left: percentage relative errors for $f(z)$ in each redshift bin for different galaxy surveys using clustering information. Right: percentage relative errors for $f(z)$ in each redshift bin for different galaxy surveys using clustering, lensing and the cross correlation information.}
  \label{Figure_mod_1}
\end{figure}
\begin{figure}[H]
\begin{center}
  	\includegraphics[width=0.52\textwidth]{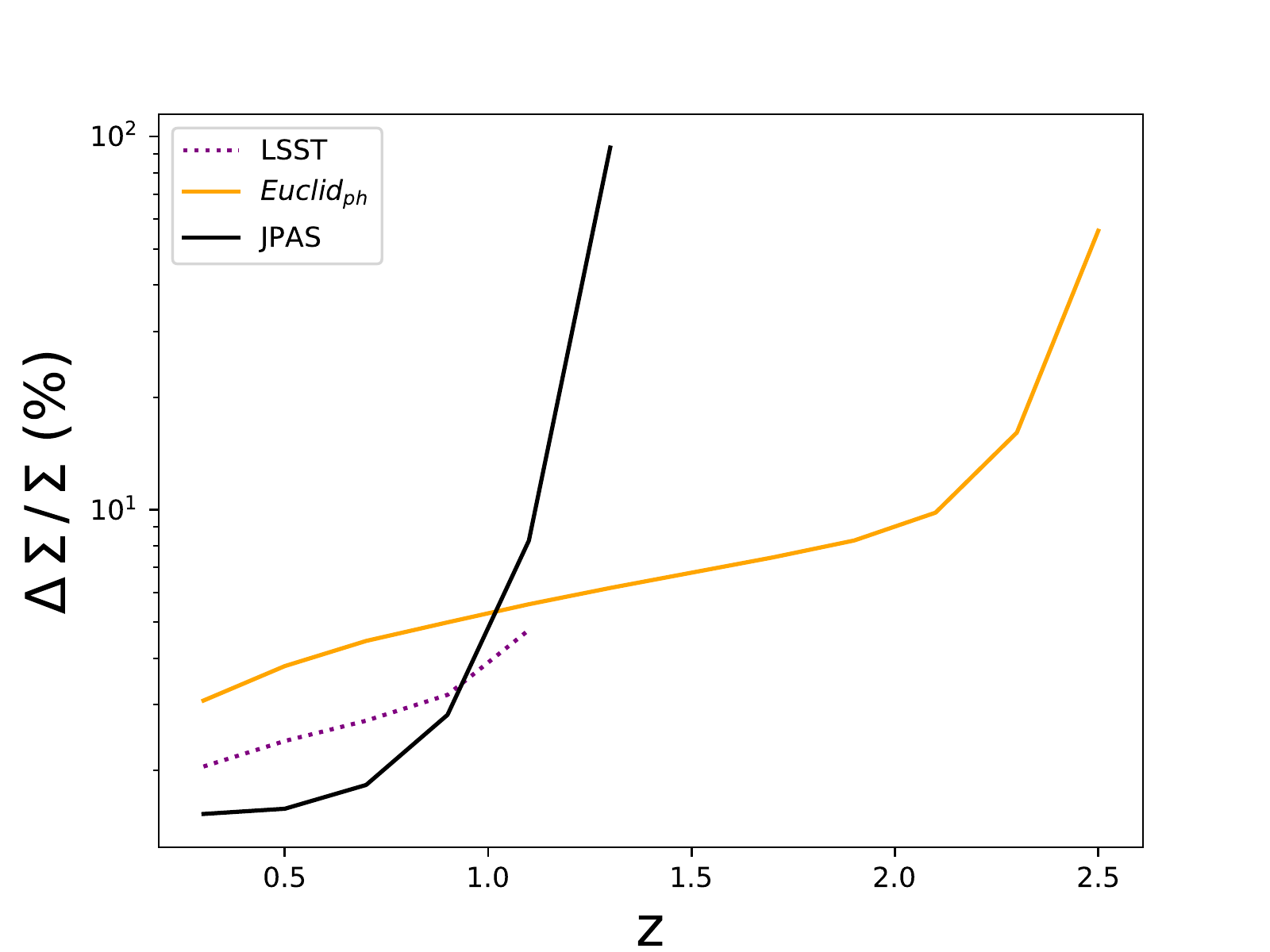}
\end{center}
		\caption{Percentage relative errors for $\Sigma(z)$ in each redshift bin for different galaxy surveys using clustering, lensing and the cross correlation information.}
  \label{Figure_mod_2}
\end{figure}

\end{widetext}

\end{document}